\documentclass{aa}  

\DeclareUnicodeCharacter{FEFF}{}
\usepackage[]{natbib}
\usepackage[T1]{fontenc}
\usepackage{ae,aecompl}
\usepackage[dvipsnames]{xcolor}
\usepackage{psfrag,color}
\usepackage[]{inputenc}
\usepackage{graphicx}	
\usepackage{amsmath}	
\usepackage{pifont}
\usepackage{amssymb}	
\usepackage{comment}
\usepackage{longtable}
\usepackage{subfigure}
\usepackage{multirow}
\usepackage{adjustbox}
\usepackage{ulem}       
\usepackage{newtxmath} 
\usepackage{newtxtext}
\usepackage{pdflscape}
\usepackage{supertabular}
\usepackage{rotating}
\usepackage{footnote}
\usepackage{times}
\usepackage{chemmacros}
\usepackage{xspace}

\usepackage[colorlinks=true, urlcolor=blue, linkcolor=blue, citecolor=blue]{hyperref} 

\makeatletter
\renewcommand*\aa@pageof{, page \thepage{} of \pageref*{LastPage}}
\makeatother

\definecolor{lime}{HTML}{A6CE39}
\DeclareRobustCommand{\orcidicon}{
	\begin{tikzpicture}
	\draw[lime, fill=lime] (0,0) 
	circle [radius=0.16] 
	node[white] {{\fontfamily{qag}\selectfont \tiny ID}};
	\draw[white, fill=white] (-0.0625,0.095) 
	circle [radius=0.007];
	\end{tikzpicture}
	\hspace{-2mm}
}

\foreach \x in {A, ..., Z}{\expandafter\xdef\csname orcid\x\endcsname{\noexpand\href{https://orcid.org/\csname orcidauthor\x\endcsname}
			{\noexpand\orcidicon}}
}

\usepackage[]{ulem}  
\newcommand\redout{\bgroup\markoverwith
{\textcolor{red}{\rule[0.5ex]{2pt}{0.8pt}}}\ULon}

\newcommand{\hi}{H\,\textsc{i}}
\newcommand{\hii}{H\,\textsc{ii}}
\newcommand{\heii}{He\,\textsc{ii}}
\newcommand{\foii}{[O\,\textsc{ii}]}
\newcommand{\foiii}{[O\,\textsc{iii}]}
\newcommand{\fnii}{[N\,\textsc{ii}]}

\newcommand{\fclii}{[Cl\,\textsc{ii}]}
\newcommand{\fcliii}{[Cl\,\textsc{iii}]}
\newcommand{\fcliv}{[Cl\,\textsc{iv}]}
\newcommand{\fsii}{[S\,\textsc{ii}]}
\newcommand{\fsiii}{[S\,\textsc{iii}]}
\newcommand{\fsiv}{[S\,\textsc{iv}]}
\newcommand{\farii}{[Ar\,\textsc{ii}]}
\newcommand{\fariii}{[Ar\,\textsc{iii}]}
\newcommand{\fariv}{[Ar\,\textsc{iv}]}
\newcommand{\ffeiii}{[Fe\,\textsc{iii}]}

\newcommand{\nel}{$n_{\rm e}$} 
\newcommand{\tel}{$T_{\rm e}$}

\begin{document}

   \title{Chlorine abundances in star-forming regions of the local Universe}

   \subtitle{}

   \author{M. Orte-Garc\'{\i}a\inst{1,2}{\orcidA{}},
          C. Esteban\inst{1,2}{\orcidB{}},
          J. E. M\'endez-Delgado
          \inst{3}{\orcidC{}},
          J. Garc\'{\i}a-Rojas\inst{1,2}{\orcidD{}}, 
          K. Z. Arellano-C\'ordova\inst{4}{\orcidE{}},
           F. F. Rosales-Ortega\inst{5}{\orcidF{}},
           E. Reyes-Rodr\'{\i}guez\inst{2,6}{\orcidG{}}
     }

   \institute{Instituto de Astrof\'isica de Canarias, E-                38205 La Laguna, Tenerife, Spain \email{maialen.orte@iac.es} 
         \and
             Departamento de Astrof\'isica, Universidad de La Laguna, E-38206 La Laguna, Tenerife, Spain 
        \and
             Instituto de Astronom\'{\i}a, Universidad Nacional Aut\'onoma de M\'exico, Ap. 70-264, 04510 CDMX, Mexico 
        \and
             Institute for Astronomy, University of Edinburgh, Royal Observatory, Edinburgh, EH9 3HJ, United Kingdom
        \and
             Instituto Nacional de Astrof\'isica, \'Optica y Electr\'onica (INAOE-CONAHCyT), Luis E. Erro 1, 72840, Tonantzintla, Puebla, Mexico
        \and
            Isaac Newton Group of Telescopes, Apto 321, E-38700 Santa Cruz de La Palma, Canary Islands, Spain
             }

   \date{Received ; accepted }

\authorrunning {Orte-Garc\'{\i}a et al.}
\titlerunning {Chlorine abundances in star-forming regions}
   \date{\today}

 \abstract
   {}
   {We study the behaviour of Cl abundance and its ratios with respect to O, S and Ar  abundances in a sample of more than 200 spectra of Galactic and extragalactic {\hii} regions and star-forming galaxies (SFGs) of the local Universe.}
   {We use the DEep Spectra of Ionised REgions Database (DESIRED) Extended project (DESIRED-E) that comprises more than 2000 spectra of {\hii} regions and SFGs with direct determinations of electron temperature ({\tel}). From this database we select those spectra where it is possible to determine the Cl$^{2+}$ abundance and whose line ratios meet certain observational criteria. We  calculate the physical conditions and Cl, O, S and Ar abundances in an homogeneous manner for all the spectra. We compare with results of photoionisation models to carry out an analysis of which is the most appropriate {\tel} indicator for the nebular volume where Cl$^{2+}$ lies, proposing a scheme that improves the determination of the Cl$^{2+}$ abundance. We compare the Cl/O ratios obtained using two different ionisation correction factor (ICF)  schemes. We also compare the nebular Cl/O distribution with stellar determinations.}
   {Our analysis indicates that the ICF scheme proposed by \citet{Izotov:06} better reproduces the observed distributions of the Cl/O ratio. We find that the log(Cl/O) vs. 12+log(O/H) and log(Cl/Ar) vs. 12+log(Ar/H) distributions are not correlated in the whole metallicity range covered by our objects indicating a lockstep evolution of those elements. In contrast, the log(Cl/S) vs. 12+log(S/H) distribution shows a weak correlation with a slight negative slope.}
   {}

   \maketitle
%

\section{Introduction}
\label{sec:introduction}

Chlorine (Cl) is an odd Z element that belongs to the group of the halogens, which also includes fluorine (F), bromine (Br) and iodine (I). 
Attending to the compilation by \citet{Asplund:21}, Cl is the 19th most abundant element in the solar photosphere. It has only two stable isotopes, $^{35}$Cl and $^{37}$Cl. These two isotopes are produced mainly during explosive oxygen burning in stars in core collapse supernova of massive stars \citep[CCSNe][]{Woosley:95}, although some contribution may be produced by Type Ia supernovae \citep[SNe Ia][]{Nomoto:97, Travaglio:04, Kobayashi:11}. $^{35}$Cl is mainly produced by proton capture on a $^{34}$S nucleus once it is formed during oxygen burning \citep{Clayton:03}, but it may also be produced from neutrino spallation in CCSNe \citep{Woosley:90}. On the other hand, $^{37}$Cl, which is 2-4 times less abundant than $^{35}$Cl \citep{Maas:18}, is produced after the radioactive decay of $^{37}$Ar, which is formed by a single neutron capture by $^{36}$Ar \citep{Clayton:03},  itself produced during the oxygen burning phase in massive stars. \citet{Prantzos:90} and \citet{Pignatari:10} propose that weak slow neutron-captures (s-process) in massive stars produces a significant amount of $^{37}$Cl. Models of Cl nucleosynthesis in CCSNe \citep{Woosley:95,  Nomoto:06, Kobayashi:11} and SNe Ia \citep{Travaglio:04} predict that the production of Cl varies as a function of the initial mass and metallicity of the stars. 

There are no atomic Cl lines observable in optical or infrared spectra of FGK stars. The only method to derive Cl abundances in stars is from spectral features of HCl molecule, which can only be measured in the $L$-band spectra of cold stars ($T_{eff} <$ 4000 K) due to the low dissociation potential of the molecule. There are Cl abundance determinations for several tens of stars, most of them M giants and a single M dwarf \citep{Maas:16, Maas:21}. In addition, the Cl abundance is difficult to estimate in diffuse and dense interstellar clouds \citep{Jenkins:09}. 

The determination of solar Cl abundance is also a complex issue, since it is not possible to obtain measurements from the quiet photosphere. \citet{Hall:72} determined the solar Cl/H ratio from HCl spectral features in sunspot spectra. \citet{Maas:16} recalculated that value with improved molecular data and new $L$-band sunspot umbrae spectra obtained by \citet{Wallace:02}, finding a value of 12+log(Cl/H)$_\odot$ = 5.31 $\pm$ 0.12. \citet{Asplund:21} argues that a larger error would be more reasonable for this quantity, proposing 12+log(Cl/H)$_\odot$ = 5.31 $\pm$ 0.20, which is the solar value they recommend. However, \citet{Maas:16} doubt the accuracy of their own determination from sunspots because it lacks radiative transfer modeling, so they adopt the meteoric values of Cl/H of 5.25 $\pm$ 0.06 obtained by \citet{Lodders:09} as the most appropriate for representing the solar Cl abundance. On the other hand, \citet{Lodders:23}, after a thorough analysis of more recent determinations of abundances of halogen elements in meteorites, recommend a present-day solar system Cl abundance of 5.23 $\pm$ 0.06. A lower value than that provided by \citet{Asplund:21}, but consistent with it considering the uncertainties. The Cl/H ratio recommended by \citet{Lodders:23} corresponds to the average abundances for a sample of carbonaceous Ivuna-type chondrites (CI-chondrites) scaled to present-day solar photospheric abundances. CI-chondrites have a composition that closely matches the solar abundances (excluding H, He and noble gases) and are expected to show halogen abundances far less contaminated than other meteorite types \citep{Lodders:23}. In fact, the Cl/H ratio find for asteroid Ryugu by \citet{Yokoyama:23} is very similar to that obtained for CI-chondrites. 

Due to the difficulties of determining Cl/H in stars, the Sun, and the neutral and molecular interstellar medium (ISM), most of the information available about cosmic Cl abundances comes from the analysis of emission lines in {\hii} regions and planetary nebulae (PNe). The {\fcliii} doublet at 5517 and 5537 \AA\ is by far the brightest emission feature of Cl in the optical spectrum of ionised nebulae. Cl$^{2+}$ is expected to be the most abundant ion of Cl in the usual ionisation conditions of {\hii} regions and PNe. However, we also expect some amount of Cl$^+$ and Cl$^{3+}$ depending on the mean ionisation degree of each nebula. This is the case, for example, of the recent work by \citet{ArellanoCordova:24}, who report the detection of a {\fcliv} line in several nearby high-ionisation degree star-forming galaxies.  Cl$^{3+}$ has an ionisation potential (IP) similar to that of He$^{+}$, so we do not expect a significant amount of Cl$^{4+}$ in {\hii} regions, where {\heii} lines are normally not detected. The brightest lines of Cl$^+$ and Cl$^{3+}$ are {\fclii} 9123 \AA\ and {\fcliv} 8046 \AA, respectively, and are very faint. For example, in the Orion Nebula they are only between 4\% and 8\% the intensity of the {\fcliii} lines (0.0002 and 0.0004 times the intensity of H$\beta$, respectively). Therefore, the detection of {\fclii} and {\fcliv} lines may only be possible in deep spectra of bright objects that also include the far red part of the optical spectrum (until about 9100 \AA). Moreover, the reddest part of the optical spectrum is heavily contaminated by intense sky absorption and emission features, so the use of high spectral resolution and/or an exquisite sky emission removal is mandatory in order to isolate those faint nebular lines. Therefore, having only {\fcliii} lines is the usual situation when trying to determine Cl abundances from spectra of {\hii} regions. In this most common situation, we must adopt an ionisation correction factor (ICF) to estimate the total Cl abundance, a factor that takes into account the fraction of the element that is in ionization states for which we do not have lines in the spectrum. In our particular case, the  ICF(Cl$^{2+}$) is a multiplicative factor to transform Cl$^{2+}$/H$^{+}$ ratios into Cl/H ones. The value of an ICF for a given nebula is usually parameterized by its mean ionisation degree, generally the  O$^{+}$/O$^{2+}$ ratio. 
For {\hii} region studies, there are different expressions for the ICF(Cl$^{2+}$) available in the literature. For example, \cite{Peimbert:77} proposed a relation 
based on the similarity of the ionisation potential of ionic species of Cl, O and S but others, as \cite{Mathis:91}, \cite{Izotov:06} or \citet{Amayo:21}, obtained fitting expressions based on photoionisation models. \citet{Esteban:15} proposed an empirical ICF(Cl$^{2+}$) relation valid for O$^{2+}$/O values from 0.14 to 1. On the other hand, \cite{DelgadoInglada:14} presented a specific ICF scheme designed for PNe. 

The radial abundance gradient of Cl in the Milky Way has been studied using PNe \citep[e.g.][]{FaundezAbans:87, Maciel:94, Henry:04} or {\hii} regions \citep{Esteban:15, ArellanoCordova:20} spectra. Of them, the most recent works \citep{Henry:04, Esteban:15, ArellanoCordova:20} give similar Cl gradient slopes and a flat Galactic Cl/O gradient, indicating a lockstep evolution of Cl and O. This is also the most common result in studies of the chemical composition of ionised gas in star-forming regions of the local Universe \citep[e.g.][]{Izotov:06, Esteban:20, Rogers:22, DominguezGuzman:22, ArellanoCordova:24}. 

In this paper, we determine the Cl abundance for a large sample of {\hii} regions and star-forming galaxies (SFGs) obtained from the literature (described in Sect.~\ref{sec:description_sample}) with good measurement of the intensity of {\fcliii} lines and direct determination of the electron temperature (\tel). In Sect.~\ref{sec:physical_ionic}, with the help of photoionisation models, we explore the most representative {\tel} diagnostics of the nebular zone where the Cl$^{2+}$ ion lies. With this prescription, we determine more accurate ionic Cl$^{2+}$ abundances. Then, in Sect.~\ref{sec:total_abundances}, we compute the total Cl abundance using different ICF(Cl$^{2+}$) schemes, as well as by directly summing the ionic abundances for those nebulae where all the ionic species of Cl are observable. In Sect.~\ref{sec:Clratios}, we explore the nucleosynthetic origin of Cl by analyzing its abundance ratio with respect to O and other elements, such as sulfur (S) or argon (Ar), which are usually assumed to evolve in lockstep with Cl. Finally, in Sect.~\ref{sec:conclusions} we summarize our main conclusions.

\section{Description of the sample}
\label{sec:description_sample}

This work is part of the DESIRED project \citep{MendezDelgado:23b}, which main aim is to carry out a homogeneous analysis of a large number of high-quality optical spectra of ionised nebulae. The  nucleus of DESIRED consists of a set of almost 200 deep intermediate- and high-spectral resolution spectra of Galactic and extragalactic \hii\ regions, SFGs, PNe and an other ionised gaseous objects observed by our research group over the last 20 years. In addition to that initial sample, we extended it including other high-quality spectra from the literature having at least one good detection of the following auroral/nebular intensity ratios of collisionally excited lines (CELs): {\foiii}~$\lambda4363/\lambda 5007$, {\fnii}~$\lambda5755/\lambda6584$, and {\fsiii}~$\lambda 6312/\lambda9069$, that can be used to derive \tel. We call this new dataset DESIRED Extended \citep[DESIRED-E,][]{MendezDelgado:24}. It currently contains 2900 spectra as of October 21, 2024. The extragalactic \hii\ regions of DESIRED-E comprise individual star-forming nebulae in spiral or irregular galaxies and the SFGs correspond to dwarf starburst galaxies whose general spectroscopic properties are indistinguishable from giant extragalactic \hii\ regions \citep[e.g.][]{Sargent:70,Melnick:85,Telles:97}. 

\begin{figure}[h!]
\centering    
\includegraphics[width=\hsize ]{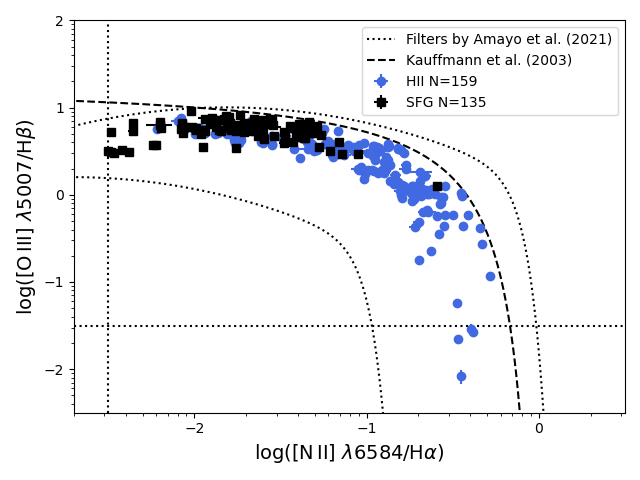}
\caption{BPT diagram of the sample of spectra of Galactic and extragalactic \hii\ regions and  star-forming galaxies compiled in DESIRED-E used in this study. The dashed line represents the empirical relation by \citet{Kauffmann:03} that we have used to distinguish between star-forming regions and active galactic nuclei (AGNs). The dotted curves and vertical and horizontal lines represent filters applied by \citet{Amayo:21} to extract photoionisation models to construct their ICF scheme for star-forming regions.  } 
\label{fig:BPT_diagram}
\end{figure}

For each spectrum included in DESIRED-E, we compile the extinction-corrected intensity ratios --and associated uncertainties-- of all the emission lines reported in the reference papers. We only consider  line intensities which quoted observational errors were less than 40\%. In this paper, we limit our study to spectra of Galactic and extragalactic \hii\ regions and SFGs of DESIRED-E for which we can determine the Cl$^{2+}$ abundance and are located below and to the left of the \citet{Kauffmann:03} line in the Baldwin-Phillips-Terlevich (BPT) diagram \citep{Baldwin:81}. In Fig.~\ref{fig:BPT_diagram}, we show the position in the BPT diagram of the objects used in this study. In Fig.~\ref{fig:BPT_diagram} we also show the filters used by \citet{Amayo:21} to select photoionisation models representative of star forming regions ({\hii} regions and SFGs) to construct their ICFs (see Sect.~\ref{sec:ionic_abundances}). The total number of DESIRED-E spectra classified as {\hii} regions or SFGs that fulfill our requirements is 294, 54.1\% corresponding to {\hii} regions and 45.9\% to SFGs. In Fig.~\ref{fig:BPT_diagram}, we can see that only four of our sample objects are slightly outside the area defined by the filters considered by\citet{Amayo:21}.

The aforementioned line intensity ratios used to derive \tel\ are insensitive to electron density, {\nel}, when it has values of the order or lower than 10$^4 \text{ cm}^{-3}$ \citep{FroeseFischer:04,Tayal:11}. The {\hii} regions and SFGs included in this study show {\nel} values well below that limit. All DESIRED-E objects have direct determinations of the total O abundance, the proxy of metallicity in ionised nebulae studies. The objects considered in this study cover a range of 12+log(O/H) values between 7.2 and 8.7 approximately. In Table A.1 we list the 294 objects included in the present study, indicating their name, whether they correspond to {\hii} regions or SFGs and the reference of their published spectra. 

\section{Physical conditions and ionic abundances}
\label{sec:physical_ionic}

We use \textit{PyNeb 1.1.18} \citep{Luridiana:15, Morisset:20} with the atomic dataset presented in Table~\ref{table:atomic_data} and the \hi~ effective recombination coefficients from \citet{Storey:95} to determine the physical conditions and the ionic abundances of the ionised nebulae. 

\begin{table*}
    \centering    
    \caption{References for atomic data used for collisionally excited lines.}
    \label{table:atomic_data}
    \begin{tabular}{ccc}
        \hline
        \noalign{\smallskip}
        Ion & Transition probabilities & Collision strengths\\
        \noalign{\smallskip}
        \hline
        \noalign{\smallskip}
            O$^{+}$   &  \citet{FroeseFischer:04} & \citet{Kisielius:09}\\
            O$^{2+}$  &  \citet{Wiese:96}, \citet{Storey:00} & \citet{Storey:14}\\
            S$^{+}$   &  \citet{Irimia:05} & \citet{Tayal:10}\\
            S$^{2+}$  &  \citet{FroeseFischer:06} & \citet{Grieve:14}\\
            Cl$^{+}$ &  \citet{Mendoza:83a} & \citet{Tayal:04}\\
            Cl$^{2+}$ &  \citet{Fritzsche:99} & \citet{Butler:89}\\
            Cl$^{3+}$ &  \citet{Mendoza:83b}; \citet{Kaufman:86}; & \citet{Butler:89}; \citet{Mendoza:83b}\\
            &  \citet{Fritzsche:99} & \\
            Ar$^{2+}$ &   \citet{Mendoza:83b}, \citet{Kaufman:86}  & \citet{Galavis:95}\\
            Ar$^{3+}$ &   \citet{Mendoza:82b}  & \citet{Ramsbottom:97}\\
            Fe$^{2+}$ & \citet{Deb:09}, \citet{Mendoza:23} & \citet{Zhang:96}, \citet{Mendoza:23}\\
        \noalign{\smallskip}
        \hline
    \end{tabular}
\end{table*}

\subsection{Electron density}
\label{subsec:densities}

We use the \textit{getCrossTemDen} routine of \textit{PyNeb} 
to derive \nel\ and \tel\ of the sample objects. This routine cross-correlates the \nel-sensitive diagnostics \fsii~$\lambda 6731/\lambda6716$, \foii~$\lambda 3726/\lambda3729$, \fcliii~$\lambda 5538/\lambda5518$, \ffeiii~$\lambda 4658/\lambda4702$, and \fariv~$\lambda 4740/\lambda4711$ with the \tel-sensitive ones \fnii~$\lambda 5755/\lambda 6584$, \foiii~$\lambda 4363/\lambda 5007$, \fariii~$\lambda 5192/\lambda 7135$, and \fsiii~$\lambda 6312/\lambda 9069$ using a Monte Carlo experiment generating 100 random values to propagate uncertainties in line intensities. This number was chosen as a compromise between calculation time and convergence of the result. As an output of this procedure, we obtain a set of \nel\ and \tel\ values along with their associated uncertainties for each diagnostic and individual point of the experiment. The average \nel\, weighted by the inverse square of the error of the 100 individual points, is adopted as the representative value of the diagnostic. We have used the Monte Carlo method for estimating the uncertainties in all physical condition and ionic abundance calculations. This is a well-suited procedure due to the complex and non-linear relationships between the various physical quantities --each one with its own uncertainty-- involved in the calculations, especially for deriving ionic abundances.

Once we obtain the \nel\ value of each diagnostic, we apply the criteria proposed by \citet{MendezDelgado:23b} to obtain a mean \nel\ representative of each nebulae. If $n_e$(\fsii~$\lambda 6731/\lambda6716) < 100\text{ cm}^{-3}$, we adopt $n_e=100 \pm 100\text{ cm}^{-3}$. If $100 \text{ cm}^{-3}\leq n_e$(\fsii~$\lambda 6731/\lambda6716)$ $< 1000\text{ cm}^{-3}$, we adopt the average between $n_e$(\fsii~$\lambda 6731/\lambda6716$) and $n_e$(\foii~$\lambda 3726/\lambda3729$). If $n_e$(\fsii~$\lambda 6731/\lambda6716)\geq 1000\text{ cm}^{-3}$, we adopt the average of $n_e$(\fsii~$\lambda 6731/\lambda6716$), $n_e$(\foii~$\lambda 3726/\lambda3729$), $n_e$(\fcliii~$\lambda 5538/\lambda5518$), $n_e$(\ffeiii~$\lambda 4658/\lambda4702$), and $n_e$(\fariv~$\lambda 4740/\lambda4711$). In cases where a value of \nel\ is not reported or could not be calculated in the source references, we adopt $n_e=100 \pm 100\text{ cm}^{-3}$. The adopted values of \nel\ for each spectra used are given in Table A.2.

\subsection{A representative electron temperature for the Cl$^{2+}$ zone}
\label{subsec:temperature}

We estimate \tel(\fnii~$\lambda 5755/\lambda 6584$), \tel(\foiii~$\lambda 4363/\lambda 5007$) and \tel(\fsiii~$\lambda 6312/\lambda 9069$) using the \textit{getTemDen} routine of \textit{PyNeb} and the aforementioned adopted average \nel\ of each object. We also use a Monte Carlo experiment of 100 random values to propagate uncertainties in \nel\ and line intensities ratios involved in the \tel\ diagnostics. To ensure a good determination of \tel\ for each object, we verify that the flux of the pairs of nebular lines coming from the same upper atomic level used (e.g., \foiii~$\lambda \lambda 5007, 4959$, \fsiii~$\lambda \lambda 9531, 9069$, \fnii~$\lambda \lambda 6584, 6548$) fit with their theoretical predictions \citep{Storey:00}. We discard any diagnostic when the observed nebular line intensity ratios differ by more than 20\% from the theoretical values. There may be deviations from the theoretical  \fsiii~$\lambda \lambda 9531, 9069$ ratio in a given object due to contamination of telluric absorption bands \citep[see][for a discussion about this issue]{MendezDelgado:24}. There are a few objects where only one of the lines of the aforementioned pairs of nebular lines is reported. In these cases, we consider the single nebular line observed in the calculation, assuming that the $T_e$ derived is valid. 

\begin{figure}[!ht]
\centering    
\includegraphics[width=\hsize ]{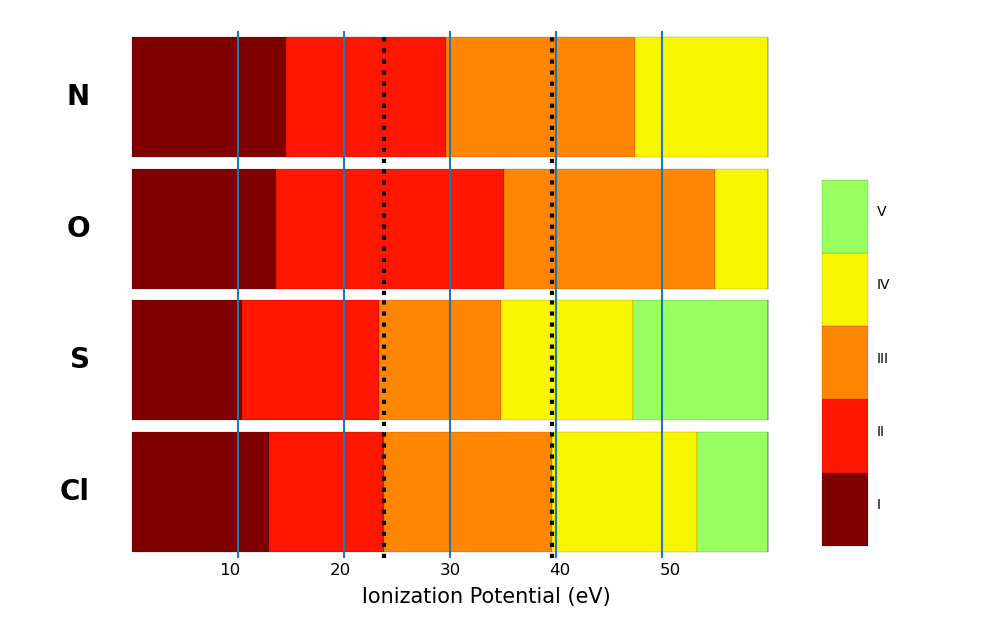}
\caption{ionisation energies of different ionic species of N, O, S and Cl we can find in usual ionisation conditions of star-forming regions. The vertical black dotted lines represent the range of photon energies capable to produce Cl$^{2+}$ ions.} 
\label{fig:IPs}
\end{figure}

\begin{figure}[!ht]
\centering    
\includegraphics[scale=0.47]{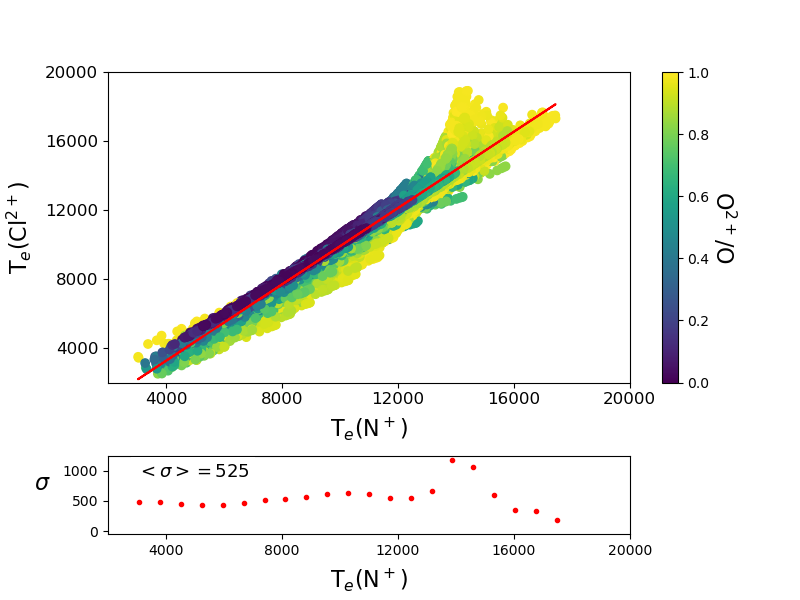}
\includegraphics[scale=0.47]{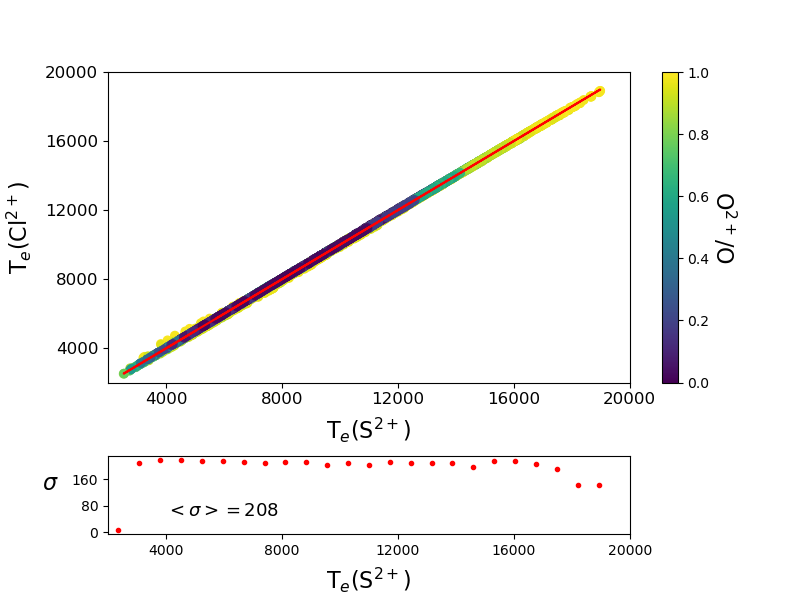}
\includegraphics[scale=0.47]{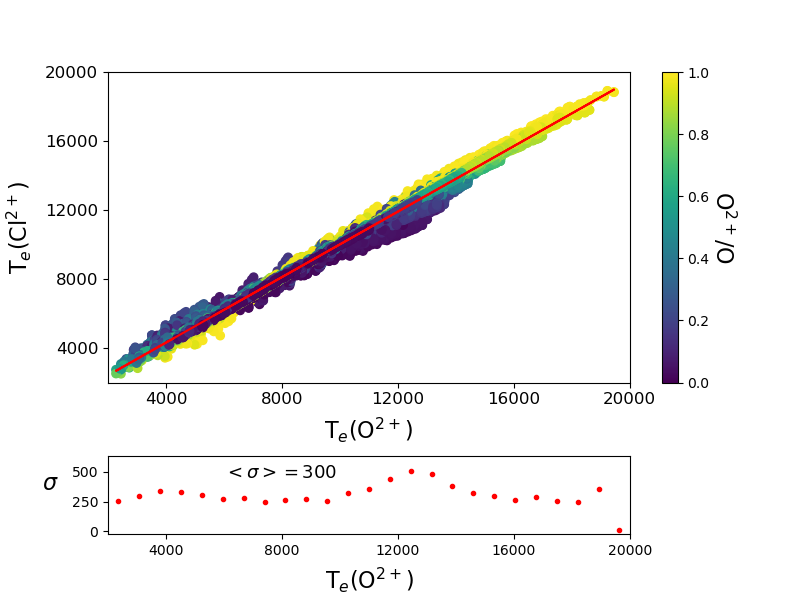}
\caption{\tel(Cl$^{2+}$) as a function of \tel(N$^{+}$) (top), \tel(S$^{2+}$) (middle) and \tel(O$^{2+}$) (bottom) obtained from the photoionisation models described in the text. Each point represents an individual model colour-coded by its mean ionisation degree, parameterized by the O$^{2+}$/O ratio. The red line is a linear fit to the points. The small panels represent the values of $\sigma$ to the fit in 1000 K bins. The mean $\sigma$ is shown inside the panels.} 
\label{fig:TCl2+}
\end{figure}

There is not a \tel\ diagnostic based on {\fcliii} lines, so we have to assume a \tel\ value representative of the nebular zone where Cl$^{2+}$ ion lies. This ion can be ionised by photons with energies between 23.81 and 39.61 eV. In Fig.~\ref{fig:IPs} we compare the ionisation ranges of the lower ionic species of Cl, N, O and S. We can see that the ionisation ranges of Cl$^{2+}$ and S$^{2+}$ (22.3 $-$ 34.8 eV) are the most similar ones, so assuming \tel(Cl$^{2+}$) $\approx$ \tel(\fsiii) seems to be a reasonable approximation. 

We have used the results from photoionisation models to check which of the \tel\ diagnostics available, \tel(\fnii), \tel(\fsiii) and \tel(\foiii) --or a combination of them-- better reproduces the behaviour of \tel(Cl$^{2+}$) in order to obtain more precise Cl$^{2+}$ abundances. To do that, we have used the models of giant \hii\ regions from the Mexican Million Models database\footnote{https://sites.google.com/site/mexicanmillionmodels/} \citep{Morisset:15} under the `BOND' reference \citep{ValeAsari:16}, built with \textit{Cloudy} v17.01 \citep{Ferland:17}. We apply the same selection criteria as \citet{Amayo:21} to select the appropriate subset of models. These criteria are: (a) starburst ages lower than 6 Myr; (b) ionisation bounded nebulae and density bounded ones selected by a cut of 70 per cent of the H$\beta$ flux; (c) a selection of realistic N/O, $U$, and O/H values \citep{ValeAsari:16}; (d) the application of {\hii} region filters on the BPT diagram defined in equation (3) of \citet[][see Fig.~\ref{fig:BPT_diagram}]{Amayo:21}. The final total number of individual models considered is 13558. 

In Fig.~\ref{fig:TCl2+} we show the distribution of \tel(Cl$^{2+}$) as a function of \tel(N$^{+}$), \tel(S$^{2+}$) and \tel(O$^{2+}$) obtained from the photoionisation models. Each point represents the result of an individual model colour-coded by its mean ionisation degree, parameterized by the O$^{2+}$/O ratio. The red continuous lines represent the linear least-squares fit to the points. Each panel has a small box below where we represent the standard deviation, $\sigma$, of the model points with respect to the fit as a function of \tel. Fig.~\ref{fig:TCl2+} clearly indicates that, attending to photoionisation models, \tel(S$^{2+}$) is the best proxy of \tel(Cl$^{2+}$) for typical spectra of \hii\ regions. In fact, the mean $\sigma$ of the \tel(Cl$^{2+}$) vs. \tel(S$^{2+}$) is the lowest of the three panels represented in Fig.~\ref{fig:TCl2+}. 

\begin{figure}[!ht]
\centering    
\includegraphics[scale=0.47]{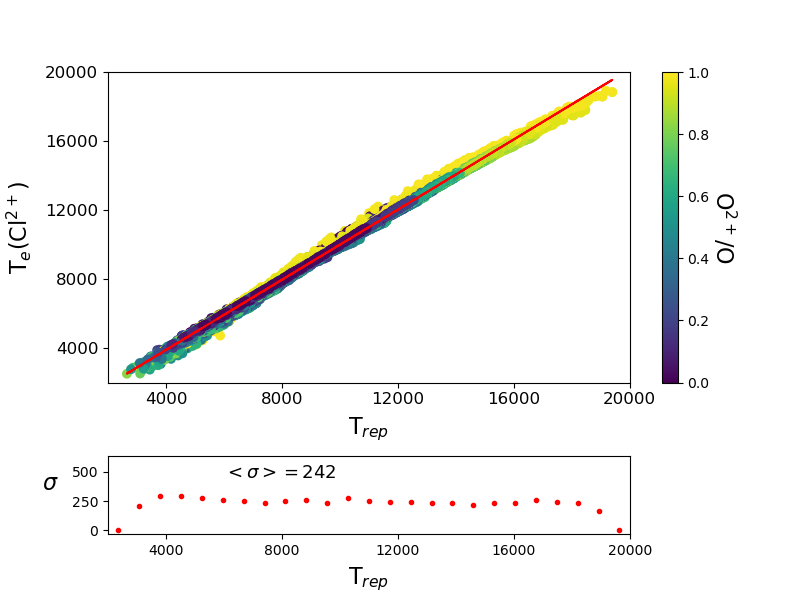}
\caption{\tel(Cl$^{2+}$) as a function of $T_{\rm rep}$ obtained from the photoionisation models described in the text. The different elements of the diagram are described in Fig.~\ref{fig:TCl2+}.} 
\label{fig:T_rep}
\end{figure}

In Table~\ref{table:No_Tes} we collect the number of objects in our DESIRED-E sample with the different combinations of the three temperature diagnostics considered, \tel(\fnii), \tel(\fsiii) and \tel(\foiii). Only 41\% of the objects have \tel(\fsiii). In order to obtain a better determination of the Cl$^{2+}$ abundance in those objects where we lack \tel(\fsiii) and following suggestions by \citet{DominguezGuzman:19}, we have defined a new \tel\ indicator more representative of \tel(Cl$^{2+}$) than \tel(\fnii) or \tel(\foiii) alone, $T_{\rm rep}$, which in the case of the models is:
\begin{equation}
\label{T_rep_mod}
    T_{\rm rep}=T_{\rm e}(\rm N^+)\cdot\left(1-\frac{O^{2+}}{O}\right)+ {\it T}_{\rm e}(\rm O^{2+})\cdot \frac{O^{2+}}{O}, 
\end{equation}
\noindent and in the case of observations is:
\begin{equation}
\label{T_re_obs}
    T_{\rm rep}=T_{\rm e}([\rm N\,\textsc{ii}])\cdot\left(1-\frac{O^{2+}}{O}\right)+ {\it T}_{\rm e}([\rm O\,\textsc{iii}])\cdot \frac{O^{2+}}{O}. 
\end{equation}
$T_{\rm rep}$ takes into account that the zone of an ionised nebula where Cl$^{2+}$ lies should be intermediate to those of N$^{+}$ and O$^{2+}$. We weight the contribution of the N$^{+}$ and O$^{2+}$ zones to $T_{\rm rep}$ using the mean O$^{2+}$/O ratio of the object. In Fig.~\ref{fig:T_rep}, we can see that the mean $\sigma$ of the model points around the fit is lower than in the \tel(Cl$^{2+}$) vs. \tel(N$^{+}$) or \tel(Cl$^{2+}$) vs. \tel(O$^{2+}$) relations separately. Table~\ref{table:No_Tes} indicates that 7.8 and 32.2 per cent of the objects have only \tel(\fnii) or \tel(\foiii) determinations available. In these cases, we have also applied $T_{\rm rep}$ but estimating the missing \tel\ indicator using the relationship between \tel(\fnii) and \tel(\foiii) proposed by \citet{Garnett:92}. The different \tel\ indicators derived for all the spectra used in this work are given in Table A.2.

\begin{table*}
\centering
    \caption{Available \tel\ determinations for the sample objects and proxies to \tel(Cl$^{2+}$) for Cl$^{2+}$/H$^+$ calculation.}
    \label{table:No_Tes}
    \begin{tabular}{ccccc}
        \hline
        \noalign{\smallskip}
        No. objects & \tel(\fnii) & \tel(\fsiii) & \tel(\foiii) & \tel(Cl$^{2+}$) proxy \\
        \noalign{\smallskip}
        \hline
        \noalign{\smallskip}
            59 & $\checkmark$ & $\checkmark$ & $\checkmark$ & \tel(\fsiii) \\
            11 & $\checkmark$ & $\checkmark$ & \ding{53} & \tel(\fsiii)  \\   
            56 & $\checkmark$ & \ding{53} & $\checkmark$ & $T_{\rm rep}$ \\
            23 & $\checkmark$ & \ding{53} & \ding{53} & $T_{\rm rep}$ with \tel(\foiii) from \citet{Garnett:92} relations \\
            47 & \ding{53} & $\checkmark$ & $\checkmark$ & \tel(\fsiii) \\
            3 & \ding{53} & $\checkmark$ & \ding{53} & \tel(\fsiii)  \\
            95 & \ding{53} & \ding{53} & $\checkmark$ & $T_{\rm rep}$ with \tel(\fnii) from \citet{Garnett:92} relations \\
        \noalign{\smallskip}
        \hline
    \end{tabular}
\end{table*}

\subsection{Ionic abundances}
\label{sec:ionic_abundances}

Once we have adopted a value of \nel\ and calculated the different \tel\ indicators for the sample spectra --including $T_{\rm rep}$ for those lacking \tel(\fsiii)-- we can proceed to calculate ionic abundances. In this work, we have determined the abundances of O$^{+}$, O$^{2+}$, S$^{+}$, S$^{2+}$, Cl$^+$, Cl$^{2+}$, Cl$^{3+}$, Ar$^{2+}$ and Ar$^{3+}$ ionic species. For O$^{+}$, we use the \foii~$\lambda \lambda 3727, 3729$ doublet, and the sum of the auroral lines \foii~$\lambda \lambda 7319, 7320, 7330, 7331$ when the bluer \foii~doublet lines are not available. In the case of O$^{2+}$, we use the sum of the bright \foiii~$\lambda \lambda 4959, 5007$ nebular lines. For S$^{+}$ and S$^{2+}$ we use the sum of \fsii~$\lambda \lambda 6716, 6731$ and \fsiii~$\lambda \lambda 9069, 9531$ doublets, respectively. There are spectra that do not cover the near-IR range, so in such cases we use the \fsiii~$\lambda 6312$ line to derive the S$^{2+}$ abundance. For Cl$^{2+}$, we use the sum of the {\fcliii} doublet at 5517 and 5537 \AA\ or one of those lines when the other is not detected. In few objects it is possible to obtain Cl$^+$ and Cl$^{3+}$ abundances, which can be  calculated from {\fclii} 9123 \AA\ and {\fcliv} 8046 \AA\ lines, respectively. In the case of Ar$^{2+}$, we use the \fariii~$\lambda 7136$ line alone or in combination with \fariii~$\lambda 7751$ when both lines are available. Finally, we use the \fariv~$\lambda \lambda 4711, 4740$ doublet to derive the Ar$^{3+}$ abundance.

We introduce \tel(\fnii) and the adopted \nel\ in the \textit{getIonAbundance} routine of \textit{PyNeb} to calculate the abundance of low-ionisation ions as O$^{+}$, S$^{+}$ and Cl$^{+}$, propagating the uncertainties through 100-point Monte Carlo experiments. When \tel(\fnii) is not available for a given spectrum, we apply the \tel\ relations of \citet{Garnett:92} to estimate it from \tel(\foiii) or \tel(\fsiii), when the first diagnostic is also  unavailable. O$^{2+}$, Cl$^{3+}$ and Ar$^{3+}$ abundances are determined using \tel(\foiii) and the adopted value of \nel. In the spectra where \tel(\foiii) can not be calculated, we use the \tel\ relations of \citet{Garnett:92} along with the values of \tel(\fnii) and/or \tel(\fsiii) when one of these are determined from the spectra. The abundance of the intermediate-ionisation ions, S$^{2+}$ and Ar$^{2+}$, is calculated using the adopted value of \nel\ and \tel(\fsiii) or \tel(\fnii) or \tel(\foiii) along with the \tel\ relations of \citet{Garnett:92} when \tel(\fsiii) is not available. Finally, as it is described in Sect.~\ref{subsec:temperature}, the Cl$^{2+}$ abundance is determined using \tel({\fsiii}) or $T_{\rm rep}$. The O$^{+}$, O$^{2+}$, S$^{+}$, S$^{2+}$, Cl$^{2+}$, Ar$^{2+}$ and Ar$^{3+}$ abundances determined for all the spectra included in this study are collected in Table A.3. The Cl$^+$ and Cl$^{3+}$ abundances in the objects where it is possible to calculate them --along to their corresponding Cl$^{2+}$ ones-- are shown separately in Table~\ref{table:Cl_abundances}.

\begin{table*}
\centering
    \caption{Ionic and total Cl abundances for objects where [Cl~{\sc ii}] and/or [Cl~{\sc iv}] lines are detected. The values are given in logarithmic units adding 12. Total Cl/H abundances are computed as the sum of the available ionic abundances or from Cl$^{2+}$/H$^+$ and using the ICF schemes of \cite{Amayo:21} (A21) or \cite{Izotov:06} (I06). The object corresponding to each ID is indicated in Table A.1.}
    \label{table:Cl_abundances}
    \begin{tabular}{lllllll}
        \hline
        \noalign{\smallskip}
        & \multicolumn{3}{c}{Cl$^{i+}$/H$^+$} & \multicolumn{3}{c}{Cl/H} \\
        ID & Cl$^{+}$ & Cl$^{2+}$ & Cl$^{3+}$ & Sum & ICF(A21) & ICF(I06)\\
        \noalign{\smallskip}
        \hline
        \noalign{\smallskip}
        \#1 & $-$ & $4.68 ^{+0.01} _{-0.02}$ & $3.07 ^{+0.08} _{-0.08}$ & $4.69 ^{+0.01} _{-0.02}$ & $4.75 ^{+0.02} _{-0.01}$ & $4.77 ^{+0.01} _{-0.02}$\\
\#4 & $3.36 ^{+0.03} _{-0.04}$ & $4.37 ^{+0.03} _{-0.04}$ & $4.40 ^{+0.02} _{-0.02}$ & $4.71 ^{+0.02} _{-0.02}$ & $4.47 ^{+0.03} _{-0.03}$ & $4.58 ^{+0.03} _{-0.04}$\\
\#5 & $-$ & $4.73 ^{+0.05} _{-0.06}$ & $3.37 ^{+0.12} _{-0.11}$ & $4.75 ^{+0.04} _{-0.05}$ & $4.79 ^{+0.05} _{-0.05}$ & $4.82 ^{+0.05} _{-0.06}$\\
\#87 & $3.60 ^{+0.06} _{-0.06}$ & $4.92 ^{+0.06} _{-0.07}$ & $3.70 ^{+0.05} _{-0.05}$ & $4.96 ^{+0.06} _{-0.07}$ & $4.99 ^{+0.06} _{-0.07}$ & $5.00 ^{+0.06} _{-0.07}$\\
\#89 & $4.21 ^{+0.02} _{-0.02}$ & $4.91 ^{+0.03} _{-0.03}$ & $-$ & $4.99 ^{+0.03} _{-0.03}$ & $5.09 ^{+0.03} _{-0.03}$ & $5.05 ^{+0.03} _{-0.03}$\\
\#90 & $3.46\pm0.03$ & $4.58 ^{+0.03} _{-0.03}$ & $2.79 ^{+0.06} _{-0.05}$ & $4.62 ^{+0.03} _{-0.03}$ & $4.65 ^{+0.03} _{-0.03}$ & $4.63 ^{+0.03} _{-0.03}$\\
\#91 & $3.80 ^{+0.05} _{-0.05}$ & $4.85 ^{+0.05} _{-0.07}$ & $3.21 ^{+0.05} _{-0.04}$ & $4.89 ^{+0.05} _{-0.06}$ & $4.92 ^{+0.06} _{-0.06}$ & $4.89 ^{+0.05} _{-0.07}$\\
\#92 & $3.16 ^{+0.05} _{-0.06}$ & $5.03 ^{+0.06} _{-0.07}$ & $3.86 ^{+0.03} _{-0.03}$ & $5.07 ^{+0.05} _{-0.07}$ & $5.17 ^{+0.07} _{-0.06}$ & $5.37 ^{+0.06} _{-0.07}$\\
\#110 & $4.30 ^{+0.04} _{-0.04}$ & $4.80 ^{+0.03} _{-0.03}$ & $-$ & $4.92 ^{+0.03} _{-0.03}$ & $4.98 ^{+0.03} _{-0.03}$ & $4.94 ^{+0.03} _{-0.03}$\\
\#111 & $-$ & $3.98 ^{+0.04} _{-0.04}$ & $2.71 ^{+0.13} _{-0.14}$ & $4.00 ^{+0.04} _{-0.04}$ & $4.13 ^{+0.04} _{-0.04}$ & $4.46 ^{+0.04} _{-0.04}$\\
\#135 & $-$ & $4.22 ^{+0.06} _{-0.06}$ & $3.59 ^{+0.05} _{-0.05}$ & $4.31 ^{+0.05} _{-0.05}$ & $4.29 ^{+0.06} _{-0.06}$ & $4.34 ^{+0.06} _{-0.06}$\\
\#139 & $3.20 ^{+0.05} _{-0.05}$ & $4.03 ^{+0.03} _{-0.04}$ & $3.27\pm0.03$ & $4.15 ^{+0.02} _{-0.03}$ & $4.10 ^{+0.03} _{-0.03}$ & $4.10 ^{+0.03} _{-0.04}$\\
\#140 & $-$ & $4.09 ^{+0.04} _{-0.04}$ & $3.32 ^{+0.03} _{-0.03}$ & $4.16 ^{+0.03} _{-0.04}$ & $4.16 ^{+0.04} _{-0.04}$ & $4.20 ^{+0.04} _{-0.04}$\\
\#141 & $-$ & $3.84 ^{+0.03} _{-0.04}$ & $3.75 ^{+0.03} _{-0.03}$ & $4.10 ^{+0.02} _{-0.02}$ & $3.97\pm0.04$ & $4.20 ^{+0.03} _{-0.04}$\\
\#211 & $-$ & $4.02 ^{+0.04} _{-0.04}$ & $3.88 ^{+0.02} _{-0.02}$ & $4.26 ^{+0.02} _{-0.02}$ & $4.10\pm0.04$ & $4.19 ^{+0.04} _{-0.04}$\\
\#239 & $-$ & $4.11 ^{+0.04} _{-0.06}$ & $3.81 ^{+0.04} _{-0.04}$ & $4.28 ^{+0.03} _{-0.04}$ & $4.18 ^{+0.05} _{-0.05}$ & $4.22 ^{+0.04} _{-0.06}$\\
\#256 & $-$ & $3.88 ^{+0.11} _{-0.12}$ & $3.83 ^{+0.07} _{-0.07}$ & $4.16 ^{+0.07} _{-0.07}$ & $4.03 ^{+0.11} _{-0.13}$ & $4.34 ^{+0.11} _{-0.12}$\\
\#262 & $-$ & $3.90 ^{+0.04} _{-0.04}$ & $3.62 ^{+0.06} _{-0.06}$ & $4.08 ^{+0.03} _{-0.03}$ & $4.04 ^{+0.04} _{-0.04}$ & $4.28 ^{+0.04} _{-0.04}$\\
\#269 & $-$ & $4.29 ^{+0.06} _{-0.08}$ & $3.74 ^{+0.06} _{-0.06}$ & $4.40 ^{+0.05} _{-0.07}$ & $4.36 ^{+0.07} _{-0.07}$ & $4.40 ^{+0.06} _{-0.08}$\\
\#294 & $-$ & $3.67 ^{+0.04} _{-0.05}$ & $3.69 ^{+0.02} _{-0.03}$ & $3.99 ^{+0.02} _{-0.03}$ & $3.78 ^{+0.04} _{-0.05}$ & $3.96 ^{+0.04} _{-0.05}$\\    
        \noalign{\smallskip}
        \hline
    \end{tabular}
\end{table*}

\section{Total abundances}
\label{sec:total_abundances}

\subsection{O, S and Ar abundances}
\label{subsec:OSAr_abundances}

The total O abundances have been calculated as the sum of the O$^{+}$/H$^{+}$ and O$^{2+}$/H$^{+}$ ratios. Spectroscopic studies of metal-poor {\hii} regions and SFGs \citep[e.g.][]{Izotov:99a,Berg:22,DominguezGuzman:22} indicate that the contribution of O$^{3+}$/H$^{+}$ can be considered negligible compared to the typical errors of the total O/H ratio. Total S abundances are determined from the sum of the S$^+$ and S$^{2+}$ ones and applying an ICF; we only consider those spectra where these two ionic abundances are available. We consider two ways to determine the total Ar abundance, when Ar$^{2+}$ lines are the only ones available in the spectrum and when we measure Ar$^{2+}$ and Ar$^{3+}$ lines, obviously the ICF used is different in each of these two cases.

\citet{Esteban:25} have carried out a study on the abundance patterns of alpha-elements --Ne, S and Ar-- based on the spectra of {\hii} regions and SFGs compiled in DESIRED-E. We follow the same methodology to derive the S/H and Ar/H ratios for the objects analysed in this paper. After a critical examination of different ICF schemes, \citet{Esteban:25} found that the ICFs of \citet{Izotov:06} best represent the behaviour of the S and Ar abundance patterns in the types of objects we are concerned with. \citet{Izotov:06} use the photoionisation models of SFGs published by \citet{Stasinska:03}, that use stellar energy distribution (SED) models from Starburst99 \citep{Leitherer:99} as ionisation sources. The ICFs by \citet{Izotov:06} are parameterized by the ionisation degree and the metallicity of the nebulae. They provide three equations depending on metallicity: `low'  (12+log(O/H)$\leq$ 7.2), `intermediate' (7.2 $<$ 12+log(O/H) $<$ 8.2) and `high' (12+log(O/H) $\geq$ 8.2) ranges, which are used to interpolate the appropriate value of the ICF using the O/H ratio calculated for each object. In the case of Ar, \citet{Izotov:06} have two different sets of equations, one when only Ar$^{2+}$ abundance is available and another when both Ar$^{2+}$ and Ar$^{3+}$ are available. The relations given by \citet{Izotov:06} for S and Ar are valid for values of O$^{2+}$/O $\geq$ 0.2. 
The total O, S, and Ar abundances determined for all the spectra included in this study are collected in Table A.4.

\subsection{Cl abundances}
\label{subsec:Cl_abundances}

As stated in Sect.~\ref{sec:introduction}, Cl$^{2+}$ is the ionisation state of Cl that presents the brightest emission lines in the optical spectrum of nebulae. Cl$^{+}$ and Cl$^{3+}$ also produce optical CELs, but are far much weaker, and they are detected only in a handful of objects. Only 8 of the spectra of {\hii} regions and SFGs with Cl$^{2+}$ abundance of DESIRED-E that we use in our analysis\footnote{Note that the total number of spectra with Cl$^{2+}$ abundances is 294, but 12 of them correspond to different positions of the Orion Nebula. From now on, for subsequent analysis we only consider the slit position observed by \citet{Esteban:04} for this object.} have determinations of Cl$^{+}$/H$^+$ and 18 of Cl$^{3+}$/H$^+$ ratios. They are listed in Table~\ref{table:Cl_abundances}. Of those objects only 6 have both Cl$^{+}$ and Cl$^{3+}$ abundances. 

For most of the objects with Cl$^{+}$ and/or Cl$^{3+}$ included in Table~\ref{table:Cl_abundances} we can derive the total Cl abundance without using an ICF(Cl)\footnote{It is interesting to comment that with the advent of the {\it James Webb Space Telescope} (JWST) and using MIRI MRS spectrograph (5--28 $\mu$m) it is possible to observe Cl$^+$ and Cl$^{3+}$ CELs in the mid-infrared (MIR, \fclii\ $\lambda$14.4 and \fcliv\ $\lambda\lambda$ 11.8, 20.3), although this spectral domain lacks Cl$^{2+}$ lines, the generally dominant ionic species in \hii\ regions. In principle, by combining optical and MIR observations of the same object we could obtain total Cl abundances and even of S and Ar without ICFs if we detect \fsiv\ $\lambda$10.5 and \farii\ $\lambda$6.9 lines. However, the combination of optical and MIR observations poses other types of problems, such as the aperture matching and relative flux calibration between both spectral ranges.}. Considering the ionisation degree of the objects and the previous results of \citet{Esteban:15}, we can obtain Cl/H as the sum of Cl$^+$/H$^+$ and Cl$^{2+}$/H$^+$ when O$^{2+}$/O $\lesssim$ 0.30, as in the case of the Galactic {\hii} regions M8 and Sh 2-311. We can derive Cl/H from the sum of Cl$^+$/H$^+$, Cl$^{2+}$/H$^+$ and Cl$^{3+}$/H$^+$ for the {\hii} regions: Orion Nebula, NGC~2579, NGC~3576 and NGC~3603 in the Milky Way, and N44C and N66A in the Large and Small Magellanic Cloud (LMC and SMC), respectively. Finally, we can obtain Cl/H from the sum of Cl$^{2+}$/H$^+$ and Cl$^{3+}$/H$^+$ in the case of the high-ionisation degree extragalactic {\hii} regions: 30 Dor, Hubble~V (NGC~6822),  NGC~1714, NGC~5408, N81 and N88A; and for the SFGs: HS1851+6933, Mrk~71, SHOC133, SBS~1420+540, Tol~1924$-$416 and W1702+18. 
All these high-ionisation objects show O$^{2+}$/O $\gtrsim$ 0.80 and we do not expect a significant fraction of Cl$^+$ in them. 
The total Cl abundances determined from the sum of ionic abundances of all the aforementioned objects are given in Table~\ref{table:Cl_abundances}. 

For the objects where only Cl$^{2+}$ abundance is available or Cl/H can not be derived from the ionic abundances observed, the Cl/O ratio has been determined applying the relation: 
\begin{equation}
\label{eq:ICFCl}
    \frac{\rm Cl}{\rm O} = \frac{\rm Cl^{2+}}{\rm O^{2+}} \cdot {\rm  ICF(Cl)}. 
\end{equation}

We have used two ICF(Cl)\footnote{Rigorously, in the form we define this ICF, it should be expressed as ICF(Cl$^{2+}$/O$^{2+}$). We abbreviate it as ICF(Cl) for simplicity.} schemes to derive the total Cl abundance. The first set is the one by \citet{Izotov:06} described in Sect~\ref{subsec:OSAr_abundances} and the second one is the scheme proposed by \citet{Amayo:21}. \citet{Esteban:25} described and analysed the suitability of these sets of ICFs in the case of Ne, S, and Ar (see footnote 2 in \citet{Esteban:25} for the convention used to name these ICFs). The ICFs by \citet{Amayo:21} are based on the same photoionisation models and selection criteria that we used to build the \tel\ relations presented in Sect.~\ref{subsec:temperature}. This ICF(Cl) can be applied for the entire range of the O$^{2+}$/O ratio. The total Cl abundance determined for all the spectra included in this study, both using the ICFs by \cite{Amayo:21} and \cite{Izotov:06}, are collected in Table A.4.

\begin{figure}[!ht]
\centering    
\includegraphics[width=\hsize ]{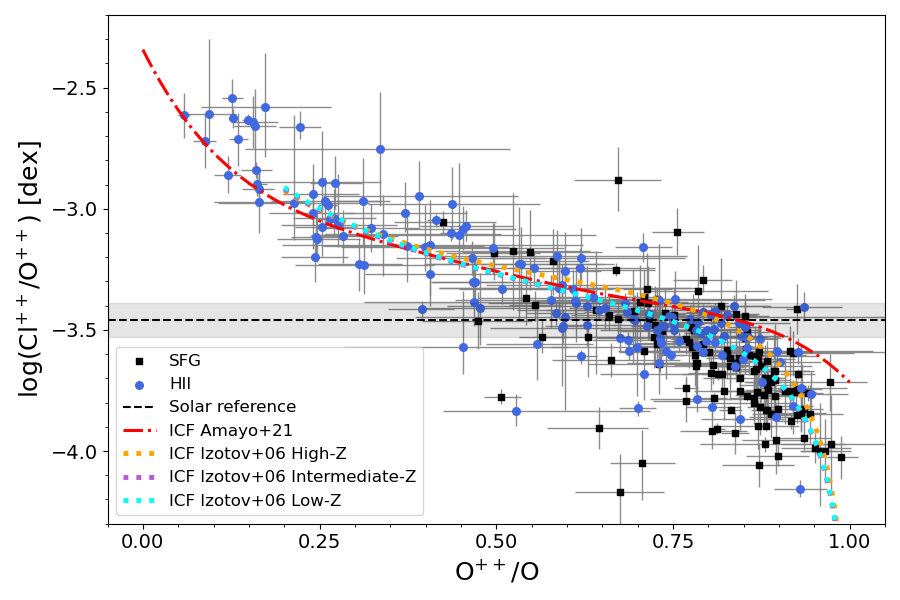}
\caption{Log(Cl$^{2+}$/O$^{2+}$) as a function of the ionisation degree, O$^{2+}$/O, for \hii\ regions (blue circles) and SFGs (black squares) of our DESIRED-E sample. The black dashed line and the grey band show the solar system log(Cl/O) and its associated uncertainty obtained from the solar system Cl/H ratio recommended by \citet{Lodders:23} and solar O/H ratio proposed by \citet{Asplund:21}. The different curves represent the ICF schemes used in this study: \citet[][red dashed-dotted lines]{Amayo:21} and \citet[][dotted lines with colors corresponding to three metallicity ranges]{Izotov:06}.} 
\label{fig:ionic_ratios_Cl}
\end{figure}

\begin{figure}[!ht]
\centering    
\includegraphics[width=\hsize ]{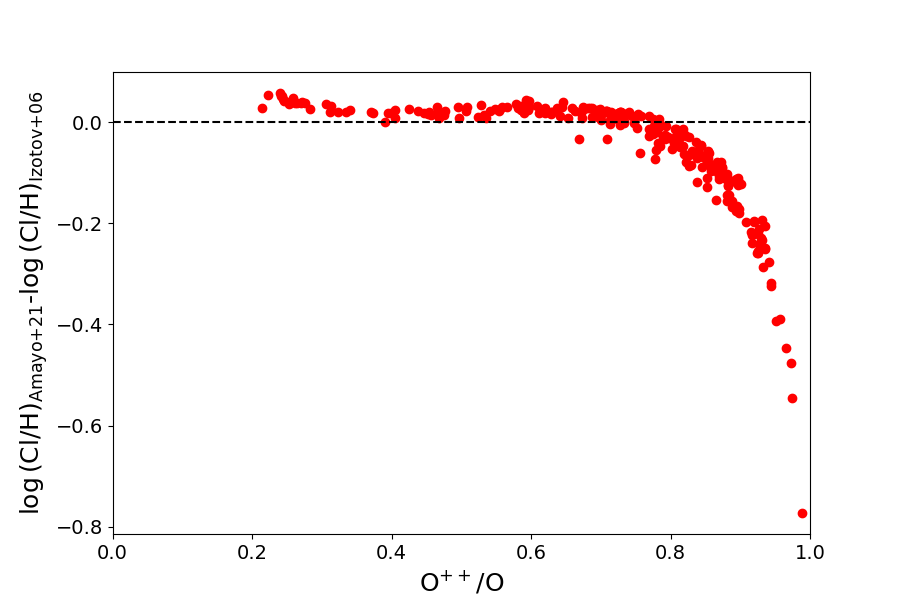}
\caption{Difference of the logarithmic Cl abundances calculated from the spectra of our DESIRED-E sample using the ICF(Cl) schemes by \citet{Amayo:21} and  \citet{Izotov:06}.} 
\label{fig:comparion_ICFs}
\end{figure}

In Fig.~\ref{fig:ionic_ratios_Cl}, we plot log(Cl$^{2+}$/O$^{2+}$) as a function of O$^{2+}$/O for the {\hii} regions and SFGs of the DESIRED-E sample. The coloured curves represent the different ICF(Cl) values defined by Eq.~\ref{eq:ICFCl} provided by the schemes considered: \citet[][red dashed-dotted lines]{Amayo:21} and \citet[][dotted lines with colours corresponding to the three metallicity ranges]{Izotov:06}. The horizontal black dashed line indicates the solar log(Cl/O) --along with its uncertainty represented by the grey band-- as reference. This solar value has been calculated using the present-day solar system Cl/H and the solar O/H ratios recommended by \citet{Lodders:23} and \citet{Asplund:21}, respectively (the reason of this choice is justified at the end of this section).  Fig.~\ref{fig:ionic_ratios_Cl} shows that the ICF(Cl) by \citet{Izotov:06} seems to reproduce better the behaviour of log(Cl$^{2+}$/O$^{2+}$) for objects with O$^{2+}$/O $>$ 0.8 (mostly SFGs). In Fig.~\ref{fig:comparion_ICFs} we show the difference of 12+log(Cl/H) calculated for all the sample objects using the ICF(Cl) schemes by \citet{Amayo:21} and \citet{Izotov:06}. The Cl abundances determined from both ICFs begin to differ significantly when O$^{2+}$/O $>$ 0.8, with those calculated with the IFC(Cl) by \citet{Izotov:06} always being higher. 

\begin{figure*}[!ht]
\centering    
\includegraphics[scale=0.40]{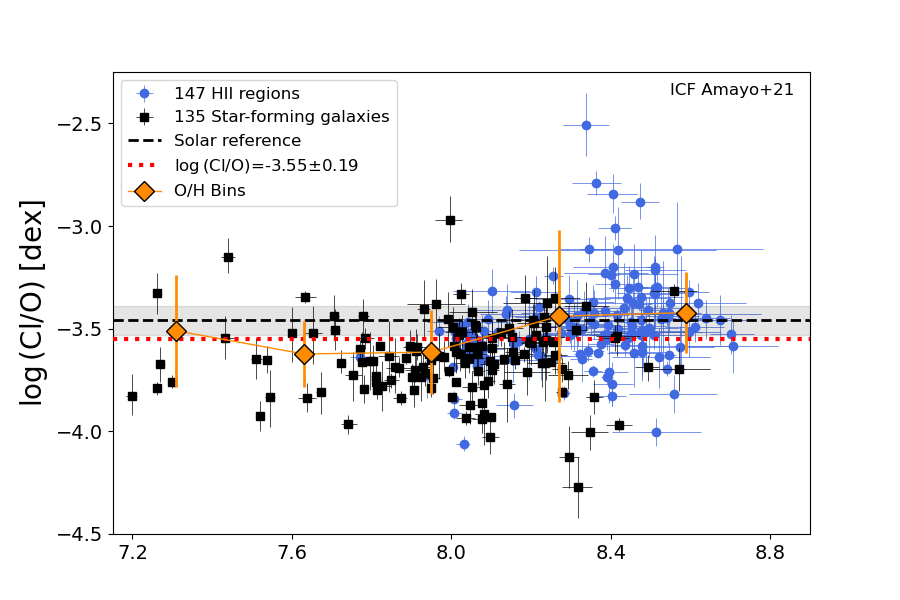}
\includegraphics[scale=0.40]{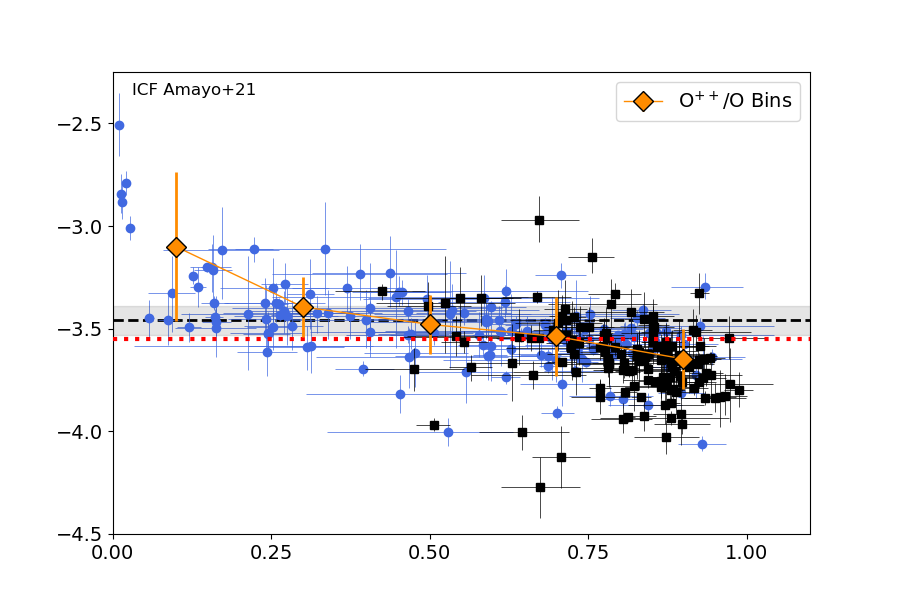}
\includegraphics[scale=0.40]{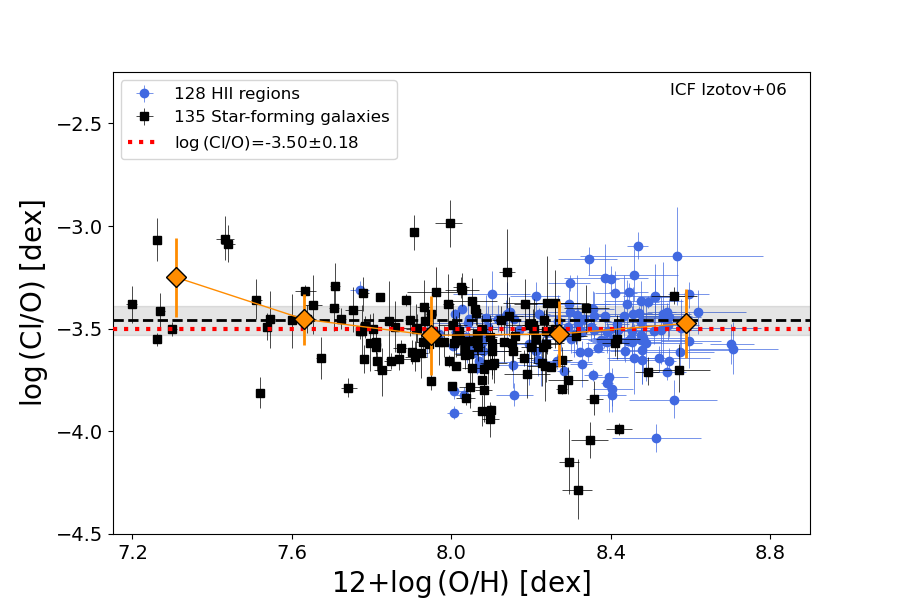}
\includegraphics[scale=0.40]{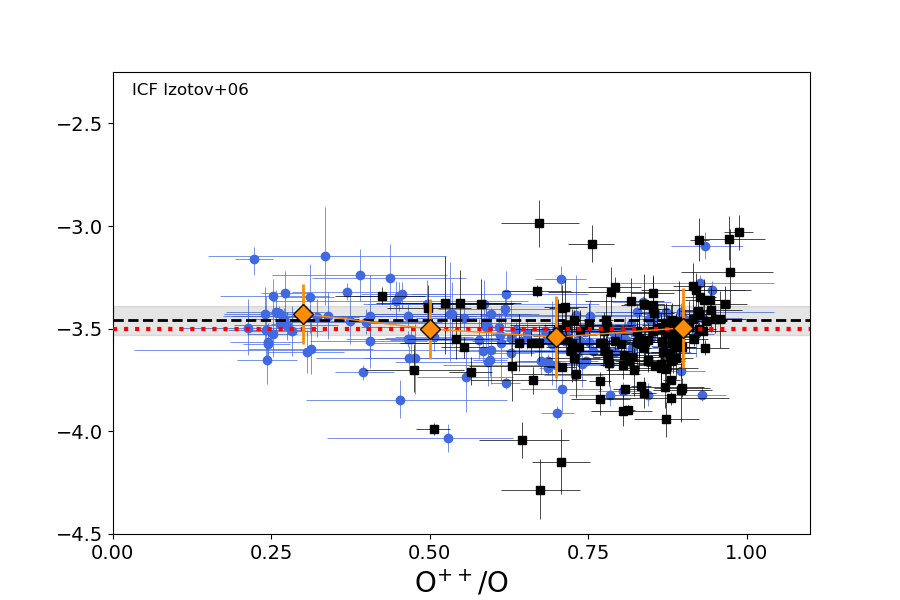}
\caption{Log(Cl/O) as a function of 12+log(O/H) (left) and the ionisation degree, O$^{2+}$/O (right), for 280 (top panels) and 261 (bottom panels)  of a sample of DESIRED-E objects. The fraction of the objects classified as \hii\ regions or SFGs each represent about 50\% of the sample. Top panels represent log(Cl/O) ratios calculated using the ICF(Cl) scheme by \citet{Amayo:21} and the bottom panels those obtained using the ICF(Cl) by \citet{Izotov:06}. The dotted red line represents the mean value of the log(Cl/O) obtained using each ICF(Cl). The orange diamonds indicate the mean log(Cl/O) values considering bins in 12+log(O/H) or O$^{2+}$/O. The black dashed lines and grey bands are obtained from the solar 12+log(O/H) and solar system log(Cl/H) ratios and their associated uncertainties recommended by \citet{Asplund:21} and \citet{Lodders:23}, respectively.} 
\label{fig:ICFCl_analysis}
\end{figure*}

In Fig.~\ref{fig:ICFCl_analysis} we show the log(Cl/O) values obtained with the two ICF(Cl) schemes used to obtain the total Cl abundances of our sample of DESIRED-E {\hii} regions and SFGs. We represent log(Cl/O) as a function of O$^{2+}$/O (right panels) and 12+log(O/H) (left panels). The number of spectra included in each panel of Fig.~\ref{fig:ICFCl_analysis} depends on the ICF(Cl) scheme used. This is because the range of O$^{2+}$/O ratios to which each scheme is applicable is different. The top panels represents the results using the ICF(Cl) by \citet{Amayo:21} and include 282\footnote{The number of objects represented in Fig.~\ref{fig:ICFCl_analysis} is lower than in Fig.~\ref{fig:ionic_ratios_Cl} because while in Fig.~\ref{fig:ionic_ratios_Cl} we include all the spectra with Cl$^{2+}$ abundance determinations, in Fig.~\ref{fig:ICFCl_analysis} we only show one of the 13 spectra from the different areas of the Orion Nebula included in DESIRED-E \citep[the one observed by][]{Esteban:04}.} spectra, 52.1\% of them correspond to {\hii} regions and 47.9\% to SFGs. The bottom panels of Fig.~\ref{fig:ICFCl_analysis} show the results obtained using the ICF(Cl) by \citet{Izotov:06}; in this case, the total number of objects represented is 263, 48.7\% of them corresponding to {\hii} regions and 51.3\% to SFGs. As in the top panels, we only include one spectrum of the Orion Nebula.

In Fig.~\ref{fig:ICFCl_analysis}, we have represented the mean value of log(Cl/O) in  four or five bins (orange diamonds), for values of 12+log(O/H) from 7.2 to 8.7 (bins 0.32 dex wide) and for O$^{2+}$/O values from 0.0 to 1.0 in the top panels and from 0.2 to 1.0 in the bottom ones (bins 0.2 wide). The top right panel indicates that the log(Cl/O) determined using the ICF(Cl) by \citet{Amayo:21} becomes lower at higher O$^{2+}$/O ratios. In fact, the mean value of log(Cl/O) of the bins decreases continuously. From the second bin (centred at O$^{2+}$/O = 0.3) to the last one on the right, the difference of the mean value of log(Cl/O) of the bins is about 0.26 dex. This trend is clearly an artifact introduced by the ICF(Cl). Considering that O/H and O$^{2+}$/O show a rough relation in {\hii} regions and SFGs of DESIRED-E \citep[see figure B.1 of][]{Esteban:25}, it is quite likely that the apparent increase of 0.21 dex in log(Cl/O) as 12+log(O/H) increases from 7.6 to 8.7 (top left panel of Fig.~\ref{fig:ICFCl_analysis}) is also actually an artifact produced by the ICF(Cl). 

In the two bottom panels of Fig.~\ref{fig:ICFCl_analysis}, we represent the log(Cl/O) values obtained using the ICF(Cl) scheme by \citet{Izotov:06}. The log(Cl/O) vs. O$^{2+}$/O distribution (bottom right panel) shows a quite flatter general behaviour. Log(Cl/O) does not show a systematic trend, and the maximum difference between the bins represented in the panel is only of about 0.11 dex. The only remarkable feature is the systematically higher log(Cl/O) values of the objects with O$^{2+}$/O $>$ 0.9. This is because the ICF(Cl) by \citet{Izotov:06} in that area varies very drastically --as can be seen in Fig.~\ref{fig:ionic_ratios_Cl}-- with the O$^{2+}$/O ratio. Any fluctuation in the Cl$^{2+}$ abundance is amplified and can produce odd Cl/H values. The log(Cl/O) vs. 12+log(O/H) distribution obtained using the ICF(Cl) by \citet{Izotov:06} is also rather flat, the data points are distributed following a slightly curved distribution with a concave shape (looking from positive ordinates). From the second bin\footnote{The first bin contains too few data points and several of them show high Cl/H ratios produced by the amplifying effect of the ICF(Cl) by \citet{Izotov:06} on objects with O$^{2+}$/O $>$ 0.9}, centred at 12+log(O/H) = 7.6, to the last one on the right, the maximum difference of the mean value of log(Cl/O) of the bins is only about 0.08 dex. Therefore, the use of the ICF(Cl) by \citet{Izotov:06} provides a rather flat log(Cl/O) distribution, whether we represent it as a function of 12+log(O/H) or O$^{2+}$/O but as long as we restrict ourselves to O$^{2+}$/O values between 0.2 and 0.9. 

\begin{figure*}[!ht]
\centering    
\includegraphics[scale=0.40]{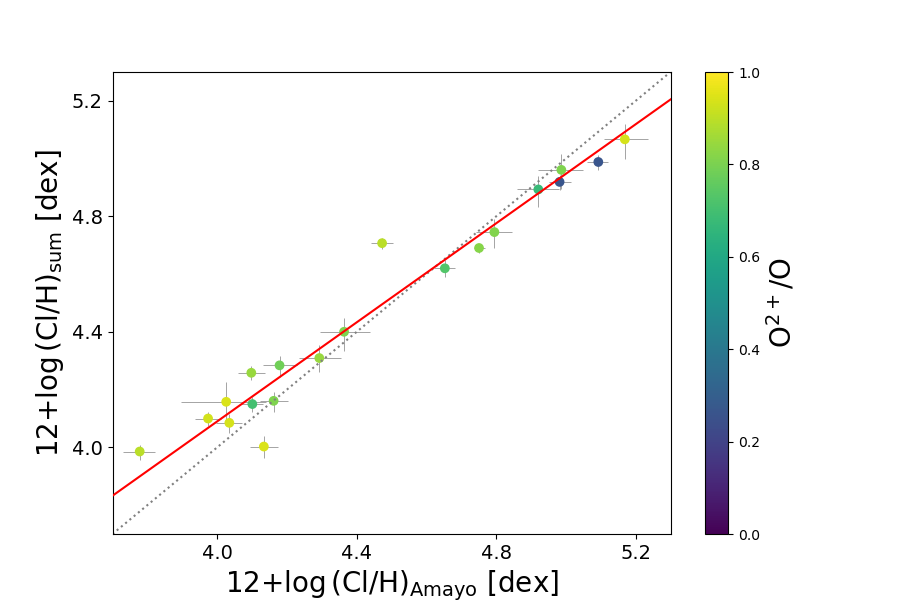}
\includegraphics[scale=0.40]{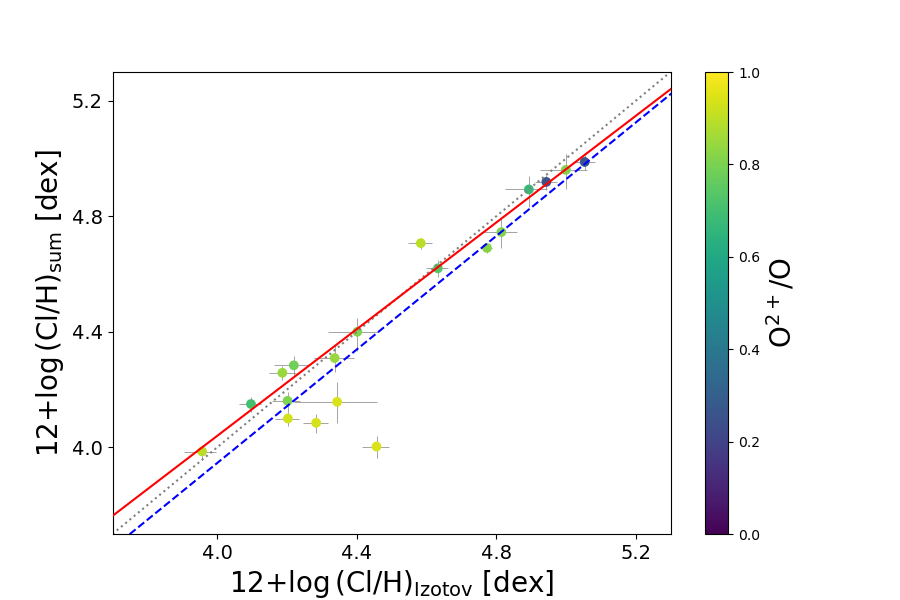}
\caption{Comparison of total Cl abundances obtained with and without ICF(Cl) for the objects where Cl$^+$ and/or Cl$^{3+}$ abundances have been derived. Left and right panels show the results obtained when the Cl/H ratio is obtained using the ICF(Cl) by \citet{Amayo:21} or \citet{Izotov:06}, respectively. Each data point is colour-coded by its ionisation degree, parameterized by the O$^{2+}$/O ratio. The dotted line indicates the 1:1 relation. The red line in left panel represents the linear fit to the data points. In the right panel, the dashed blue line represents the linear fit to all objects and the solid red line represents the fit excluding  objects with O$^{2+}$/O $>$ 0.9. } 
\label{fig:Clsuma_ClICF}
\end{figure*}

A more direct way to check which ICF(Cl) works best is to compare the Cl abundances provided by each ICF(Cl) scheme with the values obtained summing the ionic abundances of the objects included in Table~\ref{table:Cl_abundances}. In Fig.~\ref{fig:Clsuma_ClICF} we make such comparison with the data points colour-coded by its O$^{2+}$/O ratio. The diagrams show that both ICF(Cl) provide quite accurate values, with  differences not larger than 0.1 dex except for few high-ionisation degree objects. In the case of the ICF(Cl) of \citet{Amayo:21} the slope of the linear fit is 0.86; providing lower Cl/H values at low metallicities and slightly higher at higher metallicities. The ICF(Cl) of \citet{Izotov:06} gives a linear fit with a slope of 0.98 --practically parallel to the 1:1 relation-- but with a very small shift of about 0.03 dex, in the sense of obtaining slightly larger Cl/H ratios when using this ICF(Cl) and independently of the O$^{2+}$/O ratio. It is important to remark that most of the low-metallicity objects with very large O$^{2+}$/O ratios show the largest deviations with respect to the linear fit in the right panel of Fig.~\ref{fig:Clsuma_ClICF}. This is due to the steep slope of the curve describing the Izotov ICF(Cl) at O$^{2+}$/O $>$ 0.90 (see Fig.~\ref{fig:ionic_ratios_Cl}), which, as we said before, can amplify the Cl/H ratio at values that are too high. We have recalculated the linear fit in the right panel without considering the objects with O$^{2+}$/O $>$ 0.90, in this case the shift disappears and the slope becomes 0.92.  

The mean log(Cl/O) of all the data points represented in Fig.~\ref{fig:ICFCl_analysis} is $-$3.55 $\pm$ 0.19 or $-$3.50 $\pm$ 0.18 when using the ICF(Cl) by \citet{Amayo:21} or \citet{Izotov:06}, respectively. 
The solar log(Cl/O) recommended by \citep{Asplund:21} is $-3.38 \pm 0.21$ and the one obtained from the solar system Cl and O abundances recommended by \citet{Lodders:23} and \citet{Lodders:21} is $-3.50 \pm 0.09$. As can be seen, the two mean log(Cl/O) values we obtain are lower than the recommended solar one by \citet{Asplund:21} and clearly more consistent to the one obtained from the solar system data by \citet{Lodders:23} and \citet{Lodders:21}, especially the log(Cl/O) obtained using the ICF(Cl) by \citet{Izotov:06}. As discussed in Sect.~\ref{sec:introduction}, since the solar Cl/H ratio recommended by \citet{Asplund:21} comes from the sunspot determination by \citet{Maas:16}, these same authors claim that their calculation is not entirely reliable because it does not consider radiative transfer effects. We apply the same criterion as \citet{Maas:16} in assuming the meteoritic Cl/H ratio as the most representative of the solar Cl abundance. In our case the most recent determination of Cl/H = 5.23 $\pm$ 0.06 from CI-chondrites obtained by \citet{Lodders:23}. As \citet{Esteban:25} in their study on the abundance patterns of alpha-elements in the DESIRED-E sample, for the solar O/H, S/H and Ar/H ratios we adopt the values recommended by \citet{Asplund:21}. 

\section{The Cl/O, Cl/S and Cl/Ar abundance ratios in the local Universe}
\label{sec:Clratios}

According to the results obtained in Sect.~\ref{subsec:Cl_abundances}, the ICF(Cl) by \citet{Izotov:06} provides a better match with the Cl/H ratios obtained from the sum of ionic abundances, a flatter log(Cl/O) vs. O$^{2+}$/O distribution and a mean log(Cl/O) more consistent with the reference Cl/O ratio obtained from the solar system Cl abundance recommended by \citet{Lodders:23} and the O abundance recommended by \citet{Asplund:21}. We will adopt the Cl abundances set obtained using the ICF(Cl) by \citet{Izotov:06} but excluding objects with O$^{2+}$/O $>$ 0.90 throughout the discussion that follows. 

\begin{table*}
    \caption{Linear fits log(Cl/X) = $m$ $\times$ [12+log(X/H)] + $n$ per element X.}
    \label{table:fits}
    \begin{tabular}{cccccc}
        \hline
        \noalign{\smallskip}
        Ratio & No. objects & Pearson $r$ & $p$-value & Slope ($m$) & Intercept ($n$) \\
        \noalign{\smallskip}
        \hline
        \noalign{\smallskip}
        Cl/O & 237 & 0.023 & 0.729 & $0.018 \pm 0.044$ & $-3.71 \pm 0.36$ \\
        Cl/S & 229 & $-$0.209 & 0.00158 & $-0.113 \pm 0.032$ & $-1.185 \pm 0.210$ \\
        Cl/Ar & 211 & $-$0.031 & 0.655 & $-0.017 \pm 0.035$ & $-1.107 \pm 0.206$ \\
        \noalign{\smallskip}
        \hline
    \end{tabular}
\end{table*}

\begin{figure}[!ht]
\centering    
\includegraphics[width=\hsize]{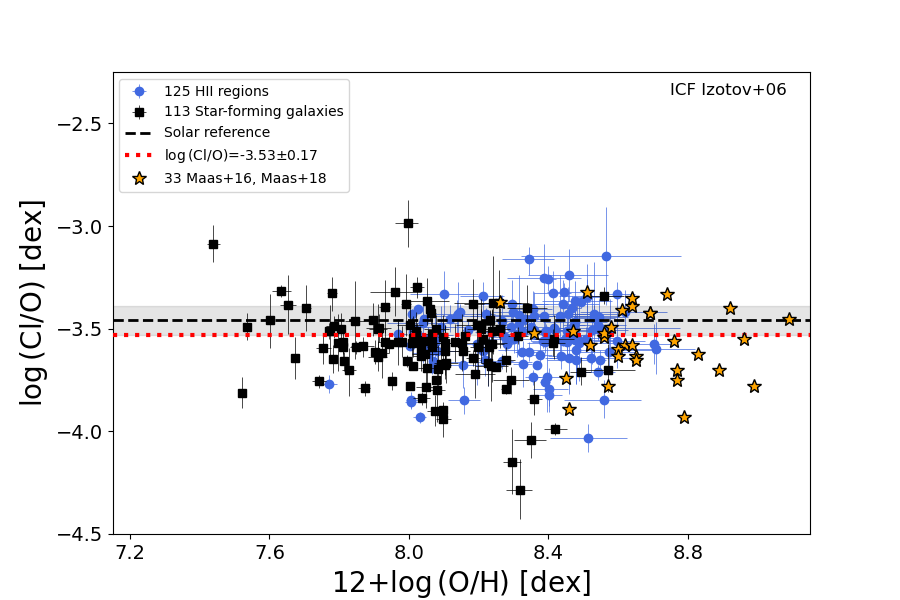}
\caption{log(Cl/O) as a function of 12+log(O/H). Black squares and blue circles represent SFGs and {\hii} regions of the DESIRED-E sample, which Cl/O ratios have been calculated using the ICF(Cl) scheme by \citet{Izotov:06}. Yellow stars represent the abundances obtained by \citet{Maas:16} and \citet{Maas:21} from spectral features of HCl molecule in the L-band spectra of nearby M giant and dwarf stars. The black dashed lines and grey bands are obtained from the solar 12+log(O/H) and solar system log(Cl/H) ratios and their associated uncertainties recommended by \citet{Asplund:21} and \citet{Lodders:23}, respectively.} 
\label{fig:ClO_stars}
\end{figure}

\subsection{Cl/O}
\label{subsec:ClO}

In Fig.~\ref{fig:ClO_stars} we represent the log(Cl/O) vs. 12+log(O/H) distribution of SFGs and {\hii} regions of the DESIRED-E sample, which Cl/O ratios have been calculated using the ICF(Cl) scheme by \citet{Izotov:06}. We also include the limited number of Cl/O ratio determinations available from stellar observations. The stellar Cl abundance is determined from spectral features of the HCl molecule in $L$-band spectra of cold stars ($T_{eff} <$ 4000 K). Using this method \citet{Maas:16} and \citet{Maas:21} obtained data for several tens of M-type giants and one dwarf, which are included as yellow stars in Fig.~\ref{fig:ClO_stars}. The stars studied by \citet{Maas:16} and \citet{Maas:21} are located at distances $<$ 1 Kpc and belong to the disc population, so most of the M-type giants of their sample must correspond to low- to intermediate-mass stars with ages typically ranging from 1$-$5 Gyr \citep{Martig:16, Miglio:21}. Taking into account the prescriptions on the nucleosynthetic origin of O and Cl, the stars which abundance data are included in Fig.~\ref{fig:ClO_stars} are not expected to have altered their initial amount of O and Cl since their formation. Therefore, their abundances should be representative of that of the ISM at the time when the stars were formed. 

In Fig.~\ref{fig:ClO_stars}, we can note that an important fraction of the stars show higher O/H ratios than most metallic {\hii} regions. This is a well-known fact. Studies comparing nebular O abundances with those of their associated O and B-type stars in Galactic star-forming regions find average differences of about 0.2 dex \citep[e.g.][]{SimonDiaz:06, SimonDiaz:11, GarciaRojas:14}, with stellar abundances being higher than nebular abundances when these last ones are calculated using CELs. This offset is the so-called abundance discrepancy problem and may be related to the presence of fluctuations in the internal spatial distribution of {\tel} in the ionised nebulae \citep{Peimbert:67, GarciaRojas:07, MendezDelgado:23a}. Although temperature fluctuations can alter the abundances calculated with CELs with respect to the true ones, the abundance ratios remain practically unchanged, so we do not expect an offset between stellar and nebular log(Cl/O) distributions \citep[e.g.][]{Esteban:20, MendezDelgado:24}. The sample of M-type stars included in Fig.~\ref{fig:ClO_stars} provide a mean log(Cl/O) = $-$3.57 $\pm$ 0.16, slightly lower than the nebular value of $-$3.50 $\pm$ 0.18 but consistent considering the standard deviation of both quantities. The value of the reference Cl/O ratio we adopt for the Sun is $-$3.46 $\pm$ 0.07.

We have performed a linear fit to the log(Cl/O) vs. 12+log(O/H) distribution of the data for {\hii} regions and SFGs represented in Fig.~\ref{fig:ClO_stars}. The parameters of the fit are given in Table~\ref{table:fits}. We can see that the slope of the fit, 0.018 $\pm$ 0.044, is very small and consistent with a flat distribution considering its uncertainty. Moreover, the sample Pearson correlation coefficient of the fit is also extremely low (0.023) and its large $p$-value  (0.729) indicates the absence of correlation between the variables involved in the linear fit\footnote{The $p$-value is used to evaluate the plausibility of the null hypothesis. A value of $p$ $>$ 0.05 indicates that the null hypothesis is plausible, i.e. there is not significant linear fit between the variables. On the other hand, when $p$ $\leq$ 0.05 it can be said that there is a significant correlation between both variables.}. Therefore, the Cl/O vs. O distribution defined by the DESIRED-E objects shows a definitively flat behaviour. If we separate the objects according to their type --{\hii} regions or SFGs-- we also find no clear indication of an evolution of the Cl/O ratio as their average metallicity increases. The average O abundance of the {\hii} regions included in Fig.~\ref{fig:ClO_stars} is 8.36 $\pm$ 0.18 while that of the SFGs is 8.10 $\pm$ 0.22. However, the mean log(Cl/O) for both types of objects is quite similar $-$3.51 $\pm$ 0.15 and $-$3.55 $\pm$ 0.19, respectively.

A basically flat distribution of the Cl/O ratio as a function of metallicity is the common result in previous works devoted to derive chemical abundances in {\hii} regions and SFGs. \citet{Izotov:06} presented the most extensive study until the present work, including the determination of Cl abundances for 148 SFGs. Those authors found a mean Cl/O ratio in agreement to the solar value and no significant trends in the log(Cl/O) vs. 12+log(O/H) distribution, but consistent with a linear fit with a slope of 0.120 $\pm$ 0.106. There are some other works dedicated to studying the behaviour of {\hii} regions in external galaxies but limited to a few tens objects at most --and a quite narrow O/H baseline-- that also show a Cl/O behaviour independent of metallicity and with mean Cl/O ratios similar to the solar one \citep[e.g.][]{Esteban:20,Rogers:22,DominguezGuzman:22}. Finally, the 15 local SFGs with enhanced star-formation rates studied by \citet{ArellanoCordova:24} exhibit a constant trend of Cl/O with metallicity but somewhat subsolar Cl/O ratios, with a mean value of log(Cl/O) = $-$3.62 $\pm$ 0.17. 

\citet{Maas:21} find also a correlation between the O and Cl abundances as well as constant Cl/Ca ratios --Ca is another alpha-element-- from spectral features of HCl molecule in nearby M giant and dwarf stars, indicating that Cl is made in similar nucleosynthesis sites as Ca or O. However, the stellar abundance ratios of \citet{Maas:21} show significant scatter. 

\subsection{Cl/S}
\label{subsec:ClS}

\begin{figure}[!ht]
\centering    
\includegraphics[width=\hsize]{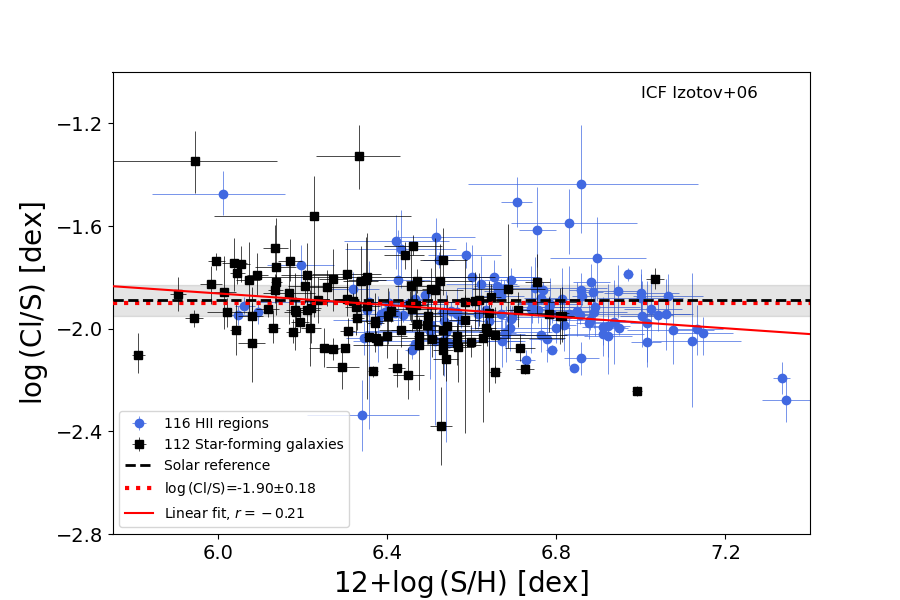}
\caption{log(Cl/S) as a function of 12+log(S/H). Black squares and blue circles represent SFGs and {\hii} regions of the DESIRED-E sample, which Cl and S abundances have been calculated using the ICF scheme by \citet{Izotov:06}. The dotted red line represents the mean log(Cl/S), while the solid one corresponds to the linear least-squares fit to the points. The black dashed lines and grey bands are obtained from the solar 12+log(S/H) and solar system log(Cl/H) ratios and their associated uncertainties recommended by \citet{Asplund:21} and \citet{Lodders:23}, respectively.} 
\label{fig:ClS}
\end{figure}

$^{35}$Cl is the most abundant stable isotope of Cl. It is mainly produced by proton capture on a $^{34}$S nucleus formed during oxygen burning \citep{Clayton:03}. In Fig.~\ref{fig:ClS} we show the log(Cl/S) vs. 12+log(S/H) distribution of 229 SFGs and {\hii} regions of the DESIRED-E sample, which Cl and S abundances have been calculated using the ICF(Cl) and ICF(S) schemes by \citet{Izotov:06}. A comprehensive analysis of the abundance patterns of S in DESIRED-E sample objects is presented in \citet{Esteban:25}. Those authors find that while the log(S/O) vs. 12+log(O/H) distribution of almost 1000 DESIRED-E spectra is rather flat, that of log(S/O) vs. 12+log(S/H) shows a linear fit with a positive slope, compatible with some contribution of S from SNe Ia, as predicted by nucleosynthesis and chemical evolution models \citep[e.g.][]{Iwamoto:99,Johnson:19,Kobayashi:20}. The linear fit to our log(Cl/S) vs. 12+log(S/H) distribution (see Table~\ref{table:fits}) indicates a rather weak correlation (Pearson $r$ coefficient = $-$0.209 and $p$ = 0.00158) with a slightly negative slope ($-$0.113 $\pm$ 0.032). The line fit is included in Fig.~\ref{fig:ClS}. The efficiency of the reaction where a $^{34}$S nucleus captures a proton to form $^{35}$Cl in stellar interiors is primarily determined on its cross section, which depends on factors such temperature and density in the stellar nucleus \citep{Lovely:21}. The weak decrease in log(Cl/S) as 12+log(S/H) increases we see in Fig.~\ref{fig:ClS} could be due to a variation in the efficiency of the proton capture reaction by $^{34}$S that produces $^{35}$Cl \citep[the dominant isotopic species of Cl, see][]{Maas:18} depending on the stellar metallicity. The chemical composition, in addition to varying the number of available nuclei, can alter the opacity of the stellar gas and vary the physical conditions in the stellar interior, such as the temperature gradient \citep[e.g.][]{Heger:03}. The behavior we observe in Fig.~\ref{fig:ClS}, although very weak, is very interesting and merits further study.

The mean log(Cl/S) of the objects included in Fig.~\ref{fig:ClS} is $-$1.90 $\pm$ 0.18, in very good agreement with the reference value of $-$1.89 $\pm$ 0.07 we adopt for the Sun assuming the solar system Cl and solar S abundances recommended by \citet{Lodders:23} and \citet{Asplund:21}, respectively. Separating the objects according to their type, we find exactly the same value of log(Cl/S) in both cases: $-$1.90, with a standard deviation of 0.17 and 0.20 in the case of {\hii} regions and SFGs, respectively. 

\subsection{Cl/Ar}
\label{subsec:ClAr}

\begin{figure}[!ht]
\centering    
\includegraphics[width=\hsize]{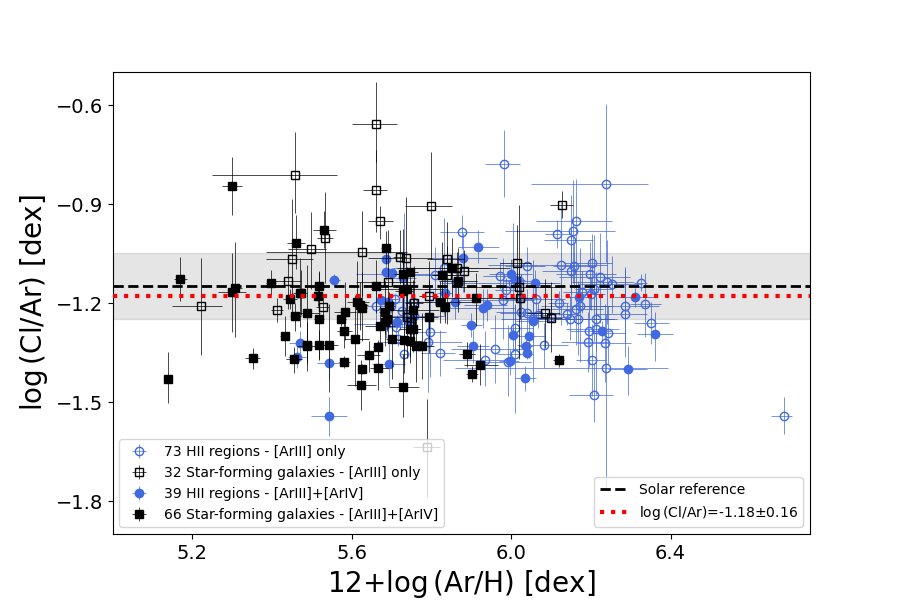}
\caption{log(Cl/Ar) as a function of 12+log(Ar/H). Black squares and blue circles represent SFGs and {\hii} regions of the DESIRED-E sample, which Cl and Ar abundances have been calculated using the ICF scheme by \citet{Izotov:06}. The dotted red line represents the mean log(Cl/Ar). The black dashed lines and grey bands are obtained from the solar 12+log(Ar/H) and solar system log(Cl/H) ratios and their associated uncertainties recommended by \citet{Asplund:21} and \citet{Lodders:23}, respectively.} 
\label{fig:ClAr}
\end{figure}

The second stable isotope of Cl is $^{37}$Cl, which is 2-4 times less abundant than $^{35}$Cl \citep{Maas:18}. It is produced after the radioactive decay of $^{37}$Ar, formed by a single neutron capture by $^{36}$Ar, which in turn is a product of the oxygen burning in massive stars \citep{Clayton:03}. Weak slow neutron-captures (s-process) in massive stars can also produce $^{37}$Cl \citep{Prantzos:90,Pignatari:10}.  In Fig.~\ref{fig:ClAr} we show the log(Cl/Ar) vs. 12+log(Ar/H) distribution of the 211 {\hii} regions and SFGs with Cl and Ar abundance determinations. Ar abundances have also been calculated using the ICF(Ar) scheme by \citet{Izotov:06}. As in the case of S, an analysis of the abundance patterns of Ar in about 1000 star-forming regions of the local Universe is presented in \citet{Esteban:25}. Those authors find that while log(Ar/O) decrease slightly as 12+log(O/H) increases, the log(Ar/O) vs. 12+log(Ar/H) relation shows a small positive slope that might be due to a small contribution of Ar from SNe Ia. The linear fit parameters given in Table~\ref{table:fits} indicate the absence of correlation between log(Cl/Ar) and 12+log(Ar/H) and that their distribution is flat. The mean log(Cl/Ar) of the {\hii} regions and SFGs included in Fig.~\ref{fig:ClAr} is $-$1.18 $\pm$ 0.16, consistent with the log(Cl/Ar) = $-$1.15 $\pm$ 0.07 we assume as representative for the Sun combining the solar system Cl abundance recommended by \citet{Lodders:23} and the solar Ar abundance proposed by \citet{Asplund:21}. Finally, the mean log(Cl/Ar) obtained for {\hii} regions is $-$1.20 $\pm$ 0.14 and $-$1.17 $\pm$ 0.18 for SFGs, indicating the lack of significant evolution of this abundance ratio with metallicity. 

\section{Conclusions}
\label{sec:conclusions}

In this paper, we analyse the behaviour of the Cl abundance and its abundance ratios with respect to O, S and Ar in more than 200 star-forming regions of the local Universe. We make use of the DEep Spectra of ionised REgions Database (DESIRED) Extended project (DESIRED-E) \citep{MendezDelgado:23b, MendezDelgado:24}, that comprises more than 2000 spectra of {\hii} regions and SFGs with direct determinations of electron temperature ({\tel}). From this database we select those spectra where it is possible to determine the Cl$^{2+}$ abundance and are located below and to the left of the \citet{Kauffmann:03} line in the Baldwin-Phillips-Terlevich (BPT) diagram \citep{Baldwin:81}. We recalculate the physical conditions ({\nel} and {\tel}) and ionic abundances of Cl$^{+}$, Cl$^{2+}$ and Cl$^{3+}$ for all spectra in an homogeneous manner. We use photoionisation models to explore the most representative {\tel} diagnostics of the nebular zone where the Cl$^{2+}$ ion lies. Our prescription for calculating more accurate Cl$^{2+}$ abundances is to use \tel(\fsiii) when it is available in the spectrum and a combination of \tel(\fnii) and \tel(\foiii) ($T_{\rm rep}$ ) when the former is not available. $T_{\rm rep}$  takes into account that the ionisation zone where Cl$^{2+}$ is found is intermediate to that occupied by  N$^{+}$ and O$^{2+}$.

We compute the total Cl abundance using the ICF(Cl) schemes by \citet{Amayo:21} and \citet{Izotov:06}, as well as simply summing the ionic abundances for those nebulae where all the expected ionic species of Cl are observable. Our analysis indicates that the ICF scheme proposed by \citet{Izotov:06} better reproduces the observed distributions of the Cl/O ratio. It produces less dependent --flatter-- distributions of the Cl/O ratios with respect to O$^{2+}$/O and 12+log(O/H) and a mean log(Cl/O) value more consistent with our assumed solar value of $-$3.46 $\pm$ 0.07, derived using the solar system Cl and solar O abundances recommended by \citet{Lodders:23} and \citet{Asplund:21}, respectively. On the other hand, the comparison of the Cl abundance obtained with and without the use of ICF(Cl) for those objects in which the Cl/H ratio can be determined from the sum of ionic abundances, shows a relationship closer to 1:1 in the case of using the ICF(Cl) scheme by \citet{Izotov:06}. Following the conclusions by \citet{Esteban:25}, we also use the ICFs schemes by \citet{Izotov:06} to calculate total S and Ar abundances. 

We compare the nebular Cl/O distribution with stellar determinations and explore the behaviour of log(Cl/O) as a function of 12+log(O/H) for 237 DESIRED-E spectra, of those the 52.7\% correspond to {\hii} regions and 47.3\% to SFGs. We find that both quantities are not correlated, indicating a lockstep evolution of both elements. We also find the same behaviour for the  log(Cl/Ar) vs. 12+log(Ar/H) distribution of our 229 DESIRED-E spectra (51.1\% HII regions and 48.9\% SFGs) for which both elemental abundances can be calculated. In contrast, the log(Cl/S) vs. 12+log(S/H) distribution of 211 DESIRED-E spectra (53.6\% HII regions and 46.4\% SFGs) shows a rather weak correlation with a slight negative slope. This behavior could be due to a variation with respect to stellar metallicity in the efficiency of the proton capture reaction by $^{34}$S that produces $^{35}$Cl, the most abundant isotopic species of Cl. 

\begin{acknowledgements}
MOG, CE, JGR and ERR acknowledge financial support from the Agencia Estatal de Investigaci\'on of the Ministerio de Ciencia e Innovaci\'on (AEI- MCINN) under grants ``The internal structure of ionised nebulae and its effects in the determination of the chemical composition of the interstellar medium and the Universe'' with reference PID2023-151648NB-I00 (DOI:10.13039/5011000110339). JGR also acknowledges financial support from the AEI-MCINN, under Severo Ochoa Centres of Excellence Programme 2020-2023 (CEX2019-000920-S), and from grant``Planetary nebulae as the key to understanding binary stellar evolution'' with reference PID-2022136653NA-I00 (DOI:10.13039/501100011033) funded by the Ministerio de Ciencia, Innovaci\'on y Universidades (MCIU/AEI) and by ERDF ``A way of making Europe'' of the European Union. JEMD  acknowledges support from project UNAM DGAPA-PAPIIT IG 101025, Mexico.\end{acknowledgements}

%
%


\onecolumn

\begin{appendix} 
\section{Tables of references, physical conditions and abundances}
\label{sec:appendix_a}

The complete reference tables associated with Appendix A can be found at: \url{https://zenodo.org/records/15005957}.

The references for the spectroscopic data are: \citet{ArellanoCordova:21}; \citet{Berg:13}; \citet{Bresolin:07a}; \citet{DelgadoInglada:16}; \citet{DominguezGuzman:22}; \citet{Esteban:04, Esteban:09, Esteban:13, Esteban:14, Esteban:17, Esteban:20}; \citet{Esteban:18}; \citet{Fernandez:18, Fernandez:22}; \citet{FernandezMartin:17}; \citet{GarciaRojas:04, GarciaRojas:05, GarciaRojas:06, GarciaRojas:07}; \citet{Guseva:00, Guseva:03c, Guseva:09, Guseva:11, Guseva:24}; \citet{Izotov:94, Izotov:97b, Izotov:04b, Izotov:06, Izotov:09, Izotov:17b, Izotov:21b}; \citet{Izotov:04}; \citet{Kurt:99}; \citet{LopezSanchez:07, Lopez-Sanchez:09}; \citet{MendezDelgado:21a, MendezDelgado:21b, MendezDelgado:22b}; \citet{MesaDelgado:09}; \citet{Noeske:00}; \citet{Peimbert:86, Peimbert:05, Peimbert:12}; \citet{Peimbert:03}; \citet{PenaGuerrero:12}; \citet{Rogers:22}; \citet{Skillman:03}; \citet{Thuan:95, Thuan:05}; \citet{Toribio:16}; \citet{Valerdi:19, Valerdi:21}; \citet{Zurita:12}

\end{appendix}
\end{document}